\documentclass[12pt]{iopart}
\usepackage{dsfont}
\usepackage{graphicx}

\newcommand{\Hilbert}{\mathcal{H}}
\newcommand{\id}{\mathds{1}}

\usepackage{iopams}
\begin{document}

\title{Markovian Master Equations: A Critical Study}

\author{\'Angel Rivas$^1$, A Douglas K Plato$^2$, Susana F Huelga$^1$ and Martin B Plenio$^{1,2}$}

\address{$^1$ Institut f\"ur Theoretische Physik, Universit\"at Ulm, Albert-Einstein-Allee 11,
D-89069 Ulm, Germany.\\
$^2$Institute for Mathematical Sciences, Imperial College London, London SW7 2PG, UK \&
QOLS, The Blackett Laboratory, Imperial College London, London SW7 2BW, UK.
}

\ead{angel.rivas@uni-ulm.de}
\begin{abstract}
We derive Markovian master equations for single and interacting harmonic systems in different scenarios, including strong internal coupling. By comparing the dynamics resulting from the corresponding master equations with numerical simulations of the global system's evolution, we delimit their validity regimes and assess the robustness of the assumptions usually made in the process of deriving the reduced Markovian dynamics. The results of these illustrative examples serve to clarify the general properties of other open quantum system scenarios subject to be treated within a Markovian approximation.
%The proposed method is sufficiently general to suggest that the conclusions made here are widely applicable to a large class of settings involving interacting chains subject to a weak interaction with an environment.
\end{abstract}

\maketitle

\section{Introduction}
It is widely assumed that one of the crucial tasks currently facing quantum
theorists is to understand and characterize the behaviour of realistic quantum systems. In any experiment, a quantum system is subject to noise and decoherence due to the unavoidable interaction with its surroundings. The theory of open quantum systems aims at developing a general framework to analyze the dynamical behaviour of systems that, as a result of their coupling with environmental degrees of freedom, will no longer evolve unitarily.
If no assumptions are made concerning the strength of the system-environment interaction and the time-correlation properties of the environment, the dynamical problem may become intractable, despite that the functional forms of very general evolutions can be derived \cite{general}. However, there exists a broad range of systems of practical interest, mostly in quantum optics and in the solid state physics, where it is possible to account for the observed dynamics
by means of a differential equation for the open system's density matrix derived in the context of Markovian processes. Such a differential equation, the so-called Markovian (or Kossakowski-Lindblad) master equation, is required to fulfill several consistency properties
such as being trace preserving and satisfying complete positivity \cite{RevKoss,BreuerPetruccione,Davies1,Davies2,Koss-Lind,Spohn,giovana,ModernCohen}.

However, from the theoretical point of view, the conditions under which these type of equations are derived are not always entirely clear, as they generally involve informal approximations motivated by a variety of microscopic models. This leaves open the range of validity of these equations, and which in some circumstances can lead to non physical evolutions. The situation becomes even worst as the complexity of the open system increases. In particular, it is not an easy question to decide whether the dynamics of a composite, possibly driven, quantum system can be described via a Markovian master equation, and if so, in what parameter regime. Actually, several groups have recently put forward operational criteria to check for deviations from Markovianity of real quantum evolutions \cite{wolf,breuer,rivas,resto}.

The main propose of this work is to study such interacting open quantum systems,
and show that there are Markovian master equations close to the real dynamics, characterizing the range of validity of each one. To this aim we have chosen a system consisting of quantum harmonic oscillators, as one can easily follow the exact dynamics using numerical simulations of a particular, but wide class
of simple states, the so-called Gaussian states. Moreover, the proposed method is general enough to be applicable to non-harmonic systems and, in particular, when the coupling between oscillators is sufficiently weak so that their local dynamics is effectively two-dimensional, we expect the conditions obtained for strict Markovianity to be directly applicable to systems of interacting qubits.

The damped harmonic oscillator is the canonical example used in most references to discuss both Markovian and non-Markovian open system dynamics (see for instance \cite{BreuerPetruccione,Haake,Gardiner,Puri,Carmichael,Weiss,Cohen} and references therein)
and exact solutions in the presence of a general environment are known \cite{HOscillator}. The dynamics of coupled damped oscillators, including those interacting with a semiclassical field, are significantly less studied, with most analysis focusing on evaluating the decoherence of initially entangled states provided that certain dynamical evolution, Markovian or not, is valid \cite{recua}. Recently, an exact master equation for two interacting harmonic oscillators subject to a global general environment was derived \cite{hu2}. Here we will focus on the derivation of Markovian master equations for interacting systems.
We will focus on a scenario where two harmonic systems are subject to independent reservoirs and present a detailed study based on the numerical simulation of the exact dynamics. The advantage of this approach is that it allows us to compute not only quantities for the damped system but also for the environment. This
enables us to check the rigour of some of the assumptions usually made in obtaining a Markovian master equation and assess their domain of validity.

We have extensively studied three damped systems. For completeness, we start our analysis by considering a single harmonic oscillator (section \ref{sectionHO}) and subsequently move to the core of our study by analyzing the dynamics of two interacting harmonic oscillators (section \ref{section2HO}), finding Markovian master equations for both weak and strong internal coupling. We finally address the
dynamics of an harmonic oscillator driven by a semiclassical field (section \ref{DrivenDampedHO}), where different Markovian master equations have been obtained and studied depending on the values of the external Rabi frequency and the detuning from the oscillator's natural frequency. To make the reading more fluent, details of the simulations and the derivation procedure are left for the appendices.

In the following two introductory sub-sections, and with the aim of setting up the notation and making the presentation as self contained as possible, we present a brief discussion of how Markovian master equations are obtained in the weak coupling limit (section \ref{sectionmaster}), and present a short review of the properties of the harmonic oscillator Gaussian states, which will be used in subsequent sections (section \ref{sectiongaussian}).

\subsection{Markovian Master Equations}\label{sectionmaster}
To derive Markovian master equations we follow the approach of projection operators initiated by Nakajima \cite{Nakajima} and Zwanzig \cite{Zwanzig}, see also \cite{BreuerPetruccione,Haake,Gardiner} for instance. In this method we define in the Hilbert space of the combined system and environment $\Hilbert=\Hilbert_S\otimes\Hilbert_E$ two orthogonal projection operators, $\mathcal{P}\rho=\Tr_E(\rho)\otimes\rho_E$ and $\mathcal{Q}=\mathds{1}-\mathcal{P}$. Here $\rho\in\mathfrak{B}(\Hilbert)$ is the combined state and $\rho_E\in\mathfrak{B}(\Hilbert_E)$ is a fixed state of the environment, which we choose to be the real initial (thermal, $k_B=1$) state,
\[
\rho_E=\rho_\mathrm{th}=\exp(-H_E/T)\{\Tr[\exp(-H_E/T)]\}^{-1}.
\]
Note that $\mathcal{P}\rho$ gives all the necessary information about the reduced system state $\rho_S$, so to know the dynamics of $\mathcal{P}\rho$ implies that one knows the time evolution of the reduced system.

We then assume that the dynamics of the whole system is given by the Hamiltonian $H=H_S+H_E+\alpha V$, where $H_S$ and $H_E$ are the individual Hamiltonians of the system and environment respectively and $V$ describes the interaction between them with coupling strength $\alpha$. Working in the interaction picture ($\hbar=1$),
\[
\tilde{\rho}(t)=\exp[i(H_S+H_E)t]\rho(t)\exp[-i(H_S+H_E)t],
\]
and analogously for $\tilde{V}(t)$, we obtain the evolution equation
\begin{equation}\label{Von-Neumann}
\frac{d}{dt}\tilde{\rho}(t)=-i\alpha[\tilde{V}(t),\tilde{\rho}(t)]\equiv\alpha\mathcal{V}(t)\tilde{\rho}(t).
\end{equation}

For the class of interactions that we are interested in $\Tr_E[\tilde{V}(t)\rho_E]=\Tr_E[\tilde{V}(t)\rho_\mathrm{th}]=0$, which implies
\begin{equation}\label{fuera1termino}
\mathcal{P}\mathcal{V}(t)\mathcal{P}=0,
\end{equation}
as can be easily checked by applying it over an arbitrary state $\rho\in\mathfrak{B}(\Hilbert)$. It is not difficult to redefine the interaction Hamiltonian such that this always holds, see for example \cite{ModernCohen,Cohen}.

Our aim is to obtain a time-evolution equation for $\mathcal{P}\rho$ under some approximation, in such a way that it describes a quantum Markovian process. To this end, we apply the projection operators on equation (\ref{Von-Neumann}),
introducing the identity $\mathds{1}=\mathcal{P}+\mathcal{Q}$ between $\mathcal{V}(t)$ and $\tilde{\rho}(t)$,
\begin{eqnarray}
\frac{d}{dt}\mathcal{P}\tilde{\rho}(t)&=\alpha\mathcal{P}\mathcal{V}(t)\mathcal{P}\tilde{\rho}(t)+\alpha\mathcal{P}\mathcal{V}(t)\mathcal{Q}\tilde{\rho}(t),\label{projectorP}\\
\frac{d}{dt}\mathcal{Q}\tilde{\rho}(t)&=\alpha\mathcal{Q}\mathcal{V}(t)\mathcal{P}\tilde{\rho}(t)+\alpha\mathcal{Q}\mathcal{V}(t)\mathcal{Q}\tilde{\rho}(t).
\end{eqnarray}
The solution of the second equation can be written formally as
\[
\mathcal{Q}\tilde{\rho}(t)=\mathcal{G}(t,t_0)\mathcal{Q}\tilde{\rho}(t_0)+\alpha\int_{t_0}^tds\mathcal{G}(t,s)\mathcal{Q}\mathcal{V}(s)\mathcal{P}\tilde{\rho}(s).
\]
This is nothing but the operational version of the variation of parameters formula for ordinary differential equations (see for example \cite{Chicone,Ince}), where the solution to the homogeneous equation
\[
\frac{d}{dt}\mathcal{Q}\tilde{\rho}(t)=\alpha\mathcal{Q}\mathcal{V}(t)\mathcal{Q}\tilde{\rho}(t)
\]
is given by the propagator
\[
\mathcal{G}(t,s)=\mathcal{T}e^{\alpha\int_s^tdt'\mathcal{Q}\mathcal{V}(t')},
\]
where $\mathcal{T}$ is the time-ordering operator. Inserting the formal solution for $\mathcal{Q}\tilde{\rho}(t)$ in (\ref{projectorP}) yields
\begin{eqnarray*}
\fl\frac{d}{dt}\mathcal{P}\tilde{\rho}(t)=\alpha\mathcal{P}\mathcal{V}(t)\mathcal{P}\tilde{\rho}(t)+\alpha\mathcal{P}\mathcal{V}(t)\mathcal{G}(t,t_0)\mathcal{Q}\tilde{\rho}(t_0)\nonumber\\
+\alpha^2\int_{t_0}^tds\mathcal{P}\mathcal{V}(t)\mathcal{G}(t,s)\mathcal{Q}\mathcal{V}(s)\mathcal{P}\tilde{\rho}(s).
\end{eqnarray*}
We now assume that the initial state of the system and bath are uncorrelated, so that the total density operator is factorised into $\rho(t_0)=\rho_S(t_0)\otimes\rho_\mathrm{th}$. From this we find $\mathcal{Q}\rho(t_0)=0$, which was guaranteed by our choice of $\mathcal{P}$ as projecting onto the initial state, and then by using (2) we finally arrive at
\begin{equation}\label{Naka-Zwan}
\frac{d}{dt}\mathcal{P}\tilde{\rho}(t)=\int_{t_0}^tds\mathcal{K}(t,s)\mathcal{P}\tilde{\rho}(s),
\end{equation}
with kernel
\[
\mathcal{K}(t,s)=\alpha^2\mathcal{P}\mathcal{V}(t)\mathcal{G}(t,s)\mathcal{Q}\mathcal{V}(s)\mathcal{P}.
\]
Equation (\ref{Naka-Zwan}) is still exact. We now consider the weak coupling limit, by taking the kernel at lowest order in $\alpha$,
\begin{equation}\label{kernel2ndorder}
\mathcal{K}(t,s)=\alpha^2\mathcal{P}\mathcal{V}(t)\mathcal{Q}\mathcal{V}(s)\mathcal{P}+\mathcal{O}(\alpha^3),
\end{equation}
so that by again using condition (\ref{fuera1termino}) we get a Born approximation for (\ref{Naka-Zwan}):
\[
\frac{d}{dt}\mathcal{P}\tilde{\rho}(t)=\alpha^2\int_{t_0}^tds\mathcal{P}\mathcal{V}(t)\mathcal{V}(s)\mathcal{P}\tilde{\rho}(s),
\]
which implies
\begin{equation}\label{Bornapprox}
\frac{d}{dt}\tilde{\rho}_S(t)=-\alpha^2\int_{t_0}^tds\Tr_E[\tilde{V}(t),[\tilde{V}(s),\tilde{\rho}_S(s)\otimes\rho_\mathrm{th}]].
\end{equation}
Note that we are not asserting here that the state of the bath is always $\rho_\mathrm{th}$, the term $\tilde{\rho}_S(s)\otimes\rho_\mathrm{th}$ appears just as a result of the application of the projection operator (see discussion in section \ref{sectionproduct}).
Now we take the initial time $t_0=0$ and an elementary change of variable $s$ by $t-s$ in the integral yields
\[
\frac{d}{dt}\tilde{\rho}_S(t)=-\alpha^2\int_0^tds\Tr_E[\tilde{V}(t),[\tilde{V}(t-s),\tilde{\rho}_S(t-s)\otimes\rho_\mathrm{th}]].
\]
We expect this equation to be valid in the limit $\alpha\rightarrow0$, but in such a limit the change in $\tilde{\rho}_S$ becomes smaller and smaller and so if we want to see dynamics we need to rescale the time by a factor $\alpha^2$ \cite{RevKoss,Davies1,Davies2} otherwise the right side of the above equation goes to zero. Thus in the limit $\alpha\rightarrow0$ the integration is extended to infinity. However in order to get a finite value for the integral, the functions $\Tr_E[\tilde{V}(t),[\tilde{V}(t-s),\rho_B]]$ must decrease appropriately. In particular this implies that they should not be periodic, which requires that the number of degrees of freedom in the environment must be infinite, as otherwise there will be a finite recurrence time.
Moreover, as $\tilde{\rho}_S$ changes very slowly in the limit $\alpha\rightarrow0$,
we can take it as a constant inside width $\tau_B$ around $s=0$ where $\Tr_E[\tilde{V}(t),[\tilde{V}(t-s),\rho_B]]$ is not zero, and so finally we obtain
\begin{equation}\label{RedfieldMarkov}
\frac{d}{dt}\tilde{\rho}_S(t)=-\alpha^2\int_0^\infty ds\Tr_E[\tilde{V}(t),[\tilde{V}(t-s),\tilde{\rho}_S(t)\otimes\rho_\mathrm{th}]].
\end{equation}
These informal arguments contain the basic ideas behind the rigorous results obtained by Davies \cite{Davies1,Davies2}.

Since we have started from a product state $\rho(t_0)=\rho_S(t_0)\otimes\rho_\mathrm{th}$, we require, for consistency, that our evolution equation generates completely positive dynamics. The last equation does not yet warrant complete positivity in the evolution \cite{Spohn}, and so we need to perform one final approximation. To this end, note that the interaction Hamiltonian may be written as:
\begin{equation}\label{Vdesc}
V=\sum_{k}A_{k}\otimes B_{k},
\end{equation}
where each $A_{k}$ can be decomposed as a sum of eigenoperators of the superoperator $[H_S,\cdot]$
\begin{equation}\label{eigenoperatorsDesc}
A_k=\sum_\nu A_k(\nu),
\end{equation}
where
\begin{equation}\label{eigenoperators}
[H_S,A_k(\nu)]=-\nu A_k(\nu).
\end{equation}
This kind of decomposition can always be made \cite{BreuerPetruccione,ModernCohen}. On the other hand, by taking the Hermitian conjugate,
\[
[H_S,A^\dagger_k(\nu)]=\nu A_k^\dagger(\nu),
\]
and since $V$ is self-adjoint, in the interaction picture one has
\[
\tilde{V}(t)=\sum_{\nu,k} e^{-i\nu t}A_k(\nu)\otimes \tilde{B}_k(t)=\sum_{\nu,k} e^{i\nu t}A_k^\dagger(\nu)\otimes \tilde{B}^\dagger_k(t).
\]
Now, substituting the decomposition in terms of $A_k(\nu)$ for $\tilde{V}(t-s)$ and $A_k^\dagger(\nu)$ for $\tilde{V}(t)$ into equation (\ref{RedfieldMarkov}) gives, after expanding the double commutator,
\begin{eqnarray}\label{beforesecular}
\fl\frac{d}{dt}\tilde{\rho}_S(t)=\sum_{\nu,\nu'}\sum_{k,\ell}e^{i(\nu'-\nu)t}\Gamma_{k,\ell}(\nu)[A_\ell(\nu)\tilde{\rho}_S(t),A_k^\dagger(\nu')]\nonumber\\
+e^{i(\nu-\nu')t}\Gamma_{\ell,k}^\ast(\nu)[A_\ell(\nu'),\tilde{\rho}_S(t)A_k^\dagger(\nu)],
\end{eqnarray}
where we have introduced the quantities
\begin{eqnarray}\label{leftFourier}
\fl\Gamma_{k,\ell}(\nu)=\alpha^2\int_0^\infty ds e^{i\nu s}\mathrm{Tr}\left[\tilde{B}^\dagger_k(t)\tilde{B}_\ell(t-s)\rho_\mathrm{th}\right]\nonumber\\
=\alpha^2\int_0^\infty ds e^{i\nu s}\mathrm{Tr}\left[\tilde{B}^\dagger_k(s) B_\ell\rho_\mathrm{th}\right],
\end{eqnarray}
with the last step being justified because $\rho_\mathrm{th}$ commutes with $\exp(iH_Et)$.

In equation (\ref{beforesecular}) the terms with different frequencies will
oscillate rapidly around zero as long as $|\nu'-\nu|\gg\alpha^2$, so in the weak coupling limit these terms vanish to obtain
\begin{equation}\label{aftersecular}
\fl\frac{d}{dt}\tilde{\rho}_S(t)=\sum_{\nu}\sum_{k,\ell}\Gamma_{k,\ell}(\nu)[A_\ell(\nu)\tilde{\rho}_S(t),A_k^\dagger(\nu)]
+\Gamma_{\ell,k}^\ast(\nu)[A_\ell(\nu),\tilde{\rho}_S(t)A_k^\dagger(\nu)].
\end{equation}
Now we decompose the matrices $\Gamma_{k,\ell}(\nu)$ as a of sum Hermitian and anti-Hermitian parts
\[
\Gamma_{k,\ell}(\nu)=\frac{1}{2}\gamma_{k,\ell}(\nu)+iS_{k,\ell}(\nu),
\]
where the coefficients
\[
S_{k,\ell}(\nu)=\frac{1}{2i}[\Gamma_{k,\ell}(\nu)-\Gamma_{\ell,k}^\ast(\nu)],
\]
and
\[
\gamma_{k,\ell}(\nu)=\Gamma_{k,\ell}(\nu)+\Gamma_{\ell,k}^\ast(\nu)=\int_{-\infty}^{\infty}dse^{i\nu s}\mathrm{Tr}\left[\tilde{B}^\dagger_k(s) B_\ell\rho_\mathrm{th}\right],
\]
form Hermitian matrices. In terms of these quantities (\ref{aftersecular}) becomes
\[
\frac{d}{dt}\tilde{\rho}_S(t)=-i[H_{\mathrm{LS}},\tilde{\rho}_S(t)]+\mathcal{D}[\tilde{\rho}_S(t)],
\]
where
\[
H_{\mathrm{LS}}=\sum_\nu\sum_{k,\ell}S_{k,\ell}A_k^\dagger(\nu)A_k(\nu),
\]
is a Hermitian operator which commutes with $H_S$, as a consequence of (\ref{eigenoperators}).
This is usually called the Shift Hamiltonian, since it produces a renormalization of the free energy levels of the system induced by the interaction with the environment. The dissipator is given by
\[
\mathcal{D}[\tilde{\rho}_S(t)]=\sum_{\nu}\sum_{k,\ell}\gamma_{k,\ell}(\nu)\left[A_\ell(\nu)\tilde{\rho}_S(t)A_k^\dagger(\nu)
-\frac{1}{2}\{A_k^\dagger(\nu)A_\ell(\nu),\tilde{\rho}_S(t)\}\right].
\]
Returning to Schr\"odinger picture, the time-evolution equation is then just
\begin{equation}\label{mastereq}
\frac{d}{dt}\rho_S(t)=-i[H_S+H_{\mathrm{LS}},\rho_S(t)]+\mathcal{D}[\rho_S(t)].
\end{equation}
Note that the matrices $\gamma_{k,\ell}(\nu)$ are positive semidefinite for every $\nu$, this is a consequence of the Bochner's theorem \cite{reedsimon1}, that is, it is easy to check that the correlation functions $\mathrm{Tr}\left[\tilde{B}^\dagger_k(s) B_\ell\rho_\mathrm{th}\right]$ are functions of positive type, and $\gamma_{k,\ell}(\nu)$ are just the Fourier transform of them. With this final remark we conclude that the equation (\ref{mastereq}) generates a completely positive semigroup \cite{Koss-Lind} and so defines a proper Markovian master equation, i.e. a completely positive semigroup.

\subsection{Gaussian States}\label{sectiongaussian}

We saw in the last section that to avoid a finite recurrence time, the number of environment degrees of freedom should strictly tend to infinity. However, in practice, the recurrence time grows very rapidly with the size of the environment and so one can still test the validity of such equations with only a finite, yet still large environment model, as long as the domain of interest is restricted to early times. The prototypical example of which is afforded by a collection on $n$ harmonic oscillators. In fact, such models are often explicitly included in master equation derivations both due to their easy handling and due to realistic physical justification. Phenomenologically speaking, they correctly describe both quantum Brownian motion and the derivation of Langevin style equations from first principles \cite{Gardiner}. However, they also provide a convenient numerical testing ground as the number of variables needed to model such systems scales polynomially in the number of degrees of freedom. This is because the harmonic oscillator falls into a class of quantum states known as Gaussian states, which are entirely characterised by their first and second moments. We now review some of their basic properties \cite{EisertPlenio03}.

For any system of $n$ canonical degrees of freedom, such as $n$ harmonic oscillators, or $n$ modes of a field, we can combine the $2n$ conjugate operators corresponding to position and momentum into a convenient row vector,
\begin{equation}
R = (x_1,x_2,... ,x_n, p_1, p_2, ..., p_n)^{\mathrm{T}}.
\end{equation}
The usual canonical commutation relations (CCR) then take the form
\begin{equation}\label{ccr}
[R_k,R_l]=i \hbar \sigma_{kl},
\end{equation}
where the skew-symmetric real $2n\times 2n$ matrix $\sigma$ is called the symplectic matrix. For the choice of $R$ above, $\sigma$ is given by,
\begin{eqnarray}
\sigma=\left[ \begin{array}{cc}
 0 & \mathds{1}_n \\
-\mathds{1}_n & 0 \end{array} \right].
\end{eqnarray}
One may also choose a mode-wise ordering of the operators, $R = (x_1,p_1,...,x_n, p_n)^{\mathrm{T}}$, in which case the symplectic matrix takes on the form,
\begin{eqnarray}\label{sympmatrix}
\sigma=\bigoplus _{j=1}^n\left[ \begin{array}{cc}
 0 & 1 \\
-1 & 0 \end{array} \right].
\end{eqnarray}
Canonical transformations of the vectors $S: R\rightarrow R'$ are then the real
$2n-$dimensional matrices $S$ which preserve the kinematic relations specified by the CCR. That is, the elements transform as $ R'_a=S_{ab}R_b$, under the
restriction,
\begin{equation}
S\sigma S^{\mathrm{T}} = \sigma.
\end{equation}
This condition defines the real $2n$-dimensional symplectic group $\mathrm{Sp}(2n,\mathds{R})$. For any element $S \in \mathrm{Sp}(2n,\mathds{R})$, the transformations $-S$, $S^{\mathrm{T}}$ and $S^{-1}$ are also symplectic matrices, and the inverse can be found from $S^{-1}=\sigma S^{\mathrm{T}} \sigma^{-1}$. The phase space then adopts the structure of a symplectic vector space, where (\ref{sympmatrix}) expresses the associated symplectic form. Rather than considering unitary operators acting on density matrices in a Hilbert space, we can instead think of all the quantum dynamics taking place on the symplectic vector space. Quantum states are then represented by functions defined on phase space, the choice of which is not unique, and common examples include the Wigner function, $Q$-function and the $P$-function. Often one has a particular benefit for a given physical problem, however for our purposes we shall consider the (Wigner) characteristic function $\chi_{\rho}(\xi)$, which we define through the Weyl
operator
\begin{equation}\label{weyl}
W_{\xi}=e^{i\xi^{\mathrm{T}} \sigma R}, \qquad \xi \in \mathds{R}^{2n}
\end{equation}
as
\begin{equation}
\chi_{\rho}(\xi)=\Tr[\rho W_{\xi}].
\end{equation}
Each characteristic function uniquely determines a quantum state. These are related
through a Fourier-Weyl transform, and so the state $\rho$ can be obtained
as
\begin{equation}\label{rhoChi}
\rho=\frac{1}{(2\pi)^{2n}} \int d^{2n} \xi \chi_{\rho}(-\xi)W_{\xi}.
\end{equation}
We then define the set of Gaussian states as those with Gaussian characteristic
functions. Equivalent definitions based on other phase
space functions also exist, but for our choice we consider characteristic functions of the form,
\begin{equation}
\chi_{\rho}(\xi)=\chi_{\rho}(0) e^{-\frac{1}{4}\xi^{\mathrm{T}}\mathfrak{C} \xi + D^{\mathrm{T}}
\xi},
\end{equation}
where $\mathfrak{C}$ is a $2n \times 2n$ real matrix
and $D\in \mathds{R}^{2n}$ is a vector. Thus, a Gaussian characteristic
function, and therefore any Gaussian state, can be completely specified by
$2n^2+3n$ real parameters. The first moments give the expectation
values of the canonical coordinates $d_j=\Tr[R_j \rho]$ and are related
to $D$ by $d=\sigma^{-1}D$, while the second moments make up the covariance
matrix defined by
\begin{equation}\label{covmat}
\mathcal{C}_{j,k}=2\mathrm{Re}\, \Tr[\rho (R_j-\langle R_j \rangle_{\rho})(R_k-\langle R_k \rangle_{\rho})].
\end{equation}
These are related to $\mathfrak{C}$ by the relation $\mathfrak{C}=\sigma^{\mathrm{T}} \mathcal{C} \sigma$. It is often the case that only the entanglement properties of a given state are of interest. As the vector $d$ can be made zero by local translations in phase space, one can specify the state entirely using
the simpler relation,
\begin{equation}
\mathcal{C}_{j,k}=2\mathrm{Re}\, \Tr[\rho R_j R_k].
\end{equation}
However, in this work we shall predominantly use the relation (\ref{covmat}). Using this convention, we mention two states of particular interest; the vacuum state, and the $n$-mode thermal state. Both take on a convenient diagonal form. In case of the vacuum this is simply the identity $\mathcal{C}=\mathds{1}_{2n}$, while for the thermal state the elements are given by
\begin{equation}
\mathcal{C}_{j,k}=\delta_{jk}\left(1+\frac{2}{e^{\omega_j/T}-1}\right),
\end{equation}
where $\omega_j$ is the frequency of the $j^{\mathrm{th}}$ mode, and the
equilibrium temperature is given by $T$.

\subsubsection{Operations on Gaussian States}
We now consider Gaussian transformations. As the $ R_j$ are hermitian and irreducible, given any real symplectic
transform $S$, the Stone-Von Neumann theorem tells us there exists a unique
unitary transformation $U_S$ acting on $\mathcal{H}$ such that $U_S W_{\xi}
U_S^{\dagger} = W_{S\xi}$. Of particular interest are those operators, $U_G$, which transform Gaussian states to Gaussian states. To this end, we consider the infinitesimal generators $G$, of Gaussian unitaries $U_G=e^{-i\epsilon G}=\mathds{1}-i\epsilon G +\mathcal{O}(\epsilon^2)$. Then to preserve the (Weyl) canonical commutation relations,
the generators $G$ must have the form $G=\sum_{j,k=1}^{2n}g_{jk}(R_j R_k-R_k R_j)/2$ \cite{EisertPlenio03}. It follows that Hamiltonians quadratic in the canonical position and momentum operators (and correspondingly the creation and
annihilation operators) will be Gaussian preserving, in particular, the Hamiltonian for $n$ simple harmonic oscillators, $H=\sum^n_{j=1}\omega_j a_j^{\dagger}a_j$. It is for this reason that harmonic oscillators
provide such a useful testing ground for many body systems.

An additional, though simple, property worth highlighting is the action of the partial trace.
Using the expression for the density matrix (\ref{rhoChi}), it is straightforward to see the effect of the partial trace
operation on the characteristic function.
If we take a mode-wise ordering of the vector $ R = (R_1, R_2)$, where $R_1$ and $R_2$ split two subspaces of $n_1$ and $n_2$ conjugate variables corresponding to partitions
of the state space of $\rho$ into $\mathcal{H}=\mathcal{H}_1\otimes\mathcal{H}_2$, then the partial trace over $\mathcal{H}_2$ is given by
\begin{equation}
\Tr_2(\rho)=\frac{1}{(2\pi)^{2n_1}} \int d^{2n_1} \xi_1 \chi_{\rho}(-\xi_1)W_{\xi_1}.
\end{equation}
That is, we need only consider the characteristic function $\chi(\xi_1)$
associated to the vector $\vec R_1$. At the level of covariance matrices,
we simply discard elements corresponding to variances including
any operators in $\vec R_2$, and so the partial trace of a Gaussian state
will itself remain Gaussian.

Finally, we make some remarks regarding closeness of two Gaussian states. Given $\rho_1$ and $\rho_2$ the fidelity between them is defined as $F(\rho_1,\rho_2)=\left(\Tr\sqrt{\sqrt{\rho_1}\rho_2\sqrt{\rho_1}}\right)^2$, and is a measure of how close both quantum system are each other. Actually a distance measure can be defined as $D_B=\sqrt{1-F}$ which is essentially the same as the Bures distance \cite{Bures} $\left(D_\mathrm{Bures}^2(\rho_1,\rho_2)=2-2\sqrt{F(\rho_1,\rho_2)}\right)$. This distance will be very useful for quantifying how well the dynamics generated by a Markovian master equation approximate the real one.

In general the fidelity is quite difficult to compute, however in the case of Gaussian states Scutaru has given closed formulas in terms of the covariance matrix \cite{Scutaru}. For example, in case of one mode Gaussian states $\rho_{G1}$ and $\rho_{G2}$, with covariance matrices $\mathcal{C}^{(1)}$ and $\mathcal{C}^{(2)}$ and displacement vectors $d^{(1)}$ and $d^{(2)}$ respectively, their fidelity is given by the formula
\begin{equation}\label{fidelity}
F(\rho_{G1},\rho_{G2})=\frac{2}{\sqrt{\Lambda+\Phi}-\sqrt{\Phi}}\exp\left[-\delta^{\mathrm{T}}\left(\mathcal{C}^{(1)}+\mathcal{C}^{(2)}\right)^{-1}\delta\right],
\end{equation}
where $\Lambda=\det\left[\mathcal{C}^{(1)}+\mathcal{C}^{(2)}\right]$, $\Phi=\det\left(\mathcal{C}^{(1)}-1\right)\det\left(\mathcal{C}^{(2)}-1\right)$ and $\delta=\left(d^{(1)}-d^{(2)}\right)$.
%--------------------------------------------------------------------------------------------------------------------------------------%
%--------------------------------------------------------------------------------------------------------------------------------------%
%--------------------------------------------------------------------------------------------------------------------------------------%

\section{Damped Harmonic Oscillator}\label{sectionHO}

\begin{figure}[h]
\centering
\includegraphics[width=0.4\textwidth]{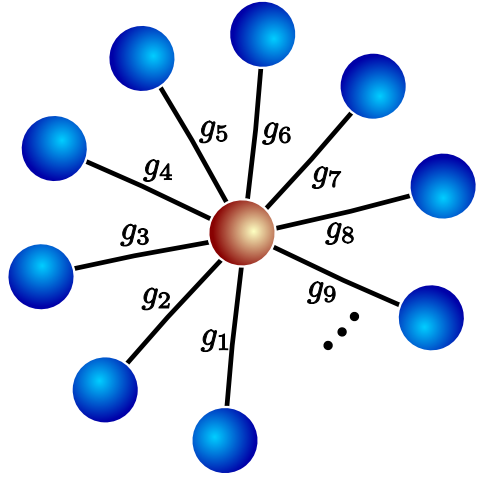}
\caption{Model for a damped harmonic oscillator. The central grey sphere represents the damped oscillator which is coupled to a large number of environmental oscillators (blue spheres) with different frequencies $\omega_j$ via the coupling constants $g_j$, These are chosen in agreement with an Ohmic spectral density (\ref{Ohmic}).}
\label{fig0}
\end{figure}

We will first consider a single harmonic oscillator damped by an environment consisting of $M$ oscillators (see figure \ref{fig0}). We want to know under which conditions the Markovian master equation that we derived in the previous section for the evolution of the damped oscillator is valid. To this aim we will approach the exact dynamical equations of the whole system when $M$ is large; these will be solved via computer simulation, and we can then compare this solution with the one obtained using a master equation.

The Hamiltonian for the whole system will be given by $(\hbar=1)$
\begin{equation}\label{SingleHam}
H=\Omega a^\dagger a+ \sum_{j=1}^M\omega_ja^\dagger_j a_j+\sum_{j=1}^Mg_j(a^\dagger a_j + a a^\dagger_j).
\end{equation}
Note that the coupling to the bath has been considered in the rotating wave approximation (RWA), which is a good description of the real dynamics for small damping $\Omega\gg\max\{g_j,j=1,\ldots,M\}$ (e.g. in the weak coupling limit) \cite{RWA}. %, which is common at optical frequencies. %\{To check the RWA is not our main issue in this work (this is a well-know topic), so for simplicity in the computation we will take it, nevertheless recall that since in the master equation derivation one assumes small $V$ everything is consistent.\}[***Consider removing or rephrasing]

For definiteness, in this paper we have chosen to distribute the environmental oscillators according to an Ohmic spectral density with exponential cut-off. In the continuous limit, this has the form \cite{Weiss}
\begin{equation}\label{Ohmic}
J(\omega)=\sum_j^Mg_j^2\delta(\omega-\omega_j)\rightarrow\alpha\omega e^{-\omega/\omega_c},
\end{equation}
where $\alpha$ is a constant which modifies the strength of the interaction and $\omega_c$ is the so-called cutoff frequency. Clearly $J(\omega)$ increases linearly for small values of $\omega$, decays exponentially for large ones, and has its maximum at $\omega=\omega_c$. Of course any other choice of spectral density could have been taken, but this in turn would require a re-analysis of the master equations' range of validity.

\subsection{Exact Solution} \label{section1exact}

The exact solution of this system can be given in terms of the time-evolution of the collection $\{a,a_j\}$ in the Heisenberg picture \cite{Puri}. From
(\ref{SingleHam}) we have
\begin{eqnarray}\label{annihi}
i\dot{a}=[a,H]=\Omega a+\sum_{j=1}^Mg_j a_j,\\
i\dot{a}_j=[a_j,H]=\omega_ja_j+g_j a,
\end{eqnarray}
and so by writing $A=(a,a_1,a_2,\ldots,a_M)^{\mathrm{T}}$, the system of differential equations may be expressed as
\begin{equation}\label{odematrix}
i\dot{A}=WA,
\end{equation}
where $W$ is the matrix
\begin{equation}\label{Wmatrix}
W=\left(\begin{array}{ccccc}
\Omega & g_1 & g_2 & \cdots & g_M\\
g_1 & \omega_1 & & & \\
g_2 & & \omega_2 & &\\
\vdots & & & \ddots &\\
g_M & & & & \omega_M
\end{array}\right),
\end{equation}
and the solution of the system will be given by
\begin{equation}
A(t)=T A(0),\quad T=e^{-iWt}.
\end{equation}
Analogously, the evolution of the creation operator will be
\begin{equation}\label{creation}
-i\dot{A}^\dagger=W A^\dagger \Rightarrow A^\dagger(t)=T^\dagger A^\dagger(0), \quad T^\dagger=e^{iWt},
\end{equation}
where $A^\dagger=(a^\dagger,a_1^\dagger,\ldots,a_M^\dagger)^{\mathrm{T}}$.

We can also compute the evolution of position and momentum operators $X=\frac{1}{2}(A+A^\dagger)$ and $P=\frac{1}{2i}(A-A^\dagger)$,
\begin{eqnarray}
\fl X(t)=\frac{1}{2}[TA(0)+T^\dagger A^\dagger(0)]\nonumber\\
=\frac{1}{2}\{T[X(0)+iP(0)] +T^\dagger [X(0)-iP(0)]\}\nonumber\\
=T_RX(0)-T_IP(0),\label{X(t)}
\end{eqnarray}
and similarly
\begin{equation}\label{Y(t)}
P(t)=T_IX(0)+T_RP(0),
\end{equation}
in these expressions, $T_R$ and $T_I$ are the self-adjoint matrices defined by
\begin{equation}\label{TRTI}
T=T_R+iT_I\Rightarrow \left\{\begin{array}{l}
T_R=\frac{T+T^\dagger}{2}=\cos(Wt)\\
T_I=\frac{T-T^\dagger}{2i}=-\sin(Wt)
\end{array}
\right. .
\end{equation}
So, the time-evolution of the vector $R=(x,x_1,\ldots,x_M,p,p_1,\ldots,p_M)^{\mathrm{T}}$ will be given by
\begin{equation}
R(t)=\mathcal{M}R(0)=\left(\begin{array}{cc}
T_R & -T_I\\
T_I & T_R
\end{array}\right)R(0),
\end{equation}
note that the size of $\mathcal{M}$ is $2(M+1)\times2(M+1)$.

Due to the linearity in the couplings in $H$, an initial (global) Gaussian state $\rho_G$ will remain Gaussian at all times $t$, and so we can restrict our attention to the evolution of its covariance matrix
\begin{equation}\label{CM}
\mathcal{C}_{i,j}=\langle R_iR_j+R_jR_i\rangle - 2\langle R_i\rangle \langle R_j\rangle.
\end{equation}
Particularly, since we are interested in just the first oscillator, we only need the evolution of the $2\times2$ submatrix $\{\mathcal{C}_{ij};i,j=1,M+2\}$. The evolution of pairs of position and momentum operators is
\begin{equation}
\langle R_i(t)R_j(t)\rangle=\sum_{k,\ell}\mathcal{M}_{i,k}\mathcal{M}_{j,\ell}\langle R_k(0)R_\ell(0)\rangle,
\end{equation}
and similarly for products of expectation values $\langle R_i(t)\rangle \langle R_j(t)\rangle$. So the elements of the covariance matrix at time $t$ will be
\begin{eqnarray*}
\fl \mathcal{C}_{i,j}(t)=\langle R_i(t)R_j(t)+R_j(t)R_i(t)\rangle - 2\langle R_i(t)\rangle \langle R_j(t)\rangle \\ =\sum_{k,\ell}\mathcal{M}_{i,k}\mathcal{M}_{j,\ell}[\langle R_k(0)R_\ell(0)+R_\ell (0) R_k(0)\rangle - 2\langle R_k(0)\rangle \langle R_\ell(0)\rangle]\\
=\sum_{k,\ell}\mathcal{M}_{i,k}\mathcal{M}_{j,\ell}\mathcal{C}_{k,\ell}(0),
\end{eqnarray*}
and for the first oscillator we have
\begin{eqnarray}
\mathcal{C}_{1,1}(t)=\sum_{k,\ell}\mathcal{M}_{1,k}\mathcal{M}_{1,\ell}\mathcal{C}_{k,\ell}(0)=(\mathcal{M}_1,\mathcal{C}\mathcal{M}_1),\label{C11}\\
\mathcal{C}_{1,M+2}(t)=\mathcal{C}_{M+2,1}(t)=\sum_{k,\ell}\mathcal{M}_{1,k}\mathcal{M}_{M+2,\ell}\mathcal{C}_{k,\ell}(0)=(\mathcal{M}_1,\mathcal{C}\mathcal{M}_{M+2}), \label{C1M+2}\\
\mathcal{C}_{M+2,M+2}(t)=\sum_{k,\ell}\mathcal{M}_{M+2,k}\mathcal{M}_{M+2,\ell}\mathcal{C}_{k,\ell}(0)=(\mathcal{M}_{M+2},\mathcal{C}\mathcal{M}_{M+2}), \label{CM+2M+2}
\end{eqnarray}
here $(\cdot,\cdot)$ denotes the scalar product, and the vectors $\mathcal{M}_1$ and $\mathcal{M}_{M+2}$ are given by
\begin{eqnarray}
\mathcal{M}_1=(\mathcal{M}_{1,1},\mathcal{M}_{1,2},\ldots,\mathcal{M}_{1,2M+2})^{\mathrm{T}},\\
\mathcal{M}_{M+2}=(\mathcal{M}_{M+2,1},\mathcal{M}_{M+2,2},\ldots,\mathcal{M}_{M+2,2M+2})^{\mathrm{T}}\label{M1M2}.
\end{eqnarray}

More details of how this exact solution is simulated in order to approach
the Markovian master equation description are given in \ref{appendixsimulation}.

\subsection{Markovian Master Equation}
The damped harmonic oscillator is a standard example for the derivation of master equations (see for example \cite{BreuerPetruccione,Puri,Carmichael,Cohen}). The Markovian master equation (\ref{mastereq}) is given by
\begin{eqnarray}\label{Master1}
\fl \frac{d}{dt}\rho(t)=-i\bar{\Omega}[a^\dagger a,\rho(t)]+\gamma(\bar{n}+1)\left(2a\rho(t) a^\dagger-a^\dagger a\rho(t)-\rho(t) a^\dagger a\right) \nonumber \\
+\gamma \bar{n}\left(2a^\dagger\rho(t) a - aa^\dagger \rho(t)-\rho(t) aa^\dagger \right),
\end{eqnarray}
where $\bar{\Omega}$ is a renormalized oscillator energy arising for the coupling to the environment
\begin{equation}\label{shift1}
\bar{\Omega}=\Omega+\Delta,\quad \Delta=\mathrm{P.V.}\int^\infty_0 d\omega \frac{J(\omega)}{\Omega-\omega},
\end{equation}
(here $\mathrm{P.V.}$ denotes the Cauchy principal value of the integral), $\bar{n}$ is the mean number of bath quanta with frequency $\Omega$, given by the Bose-Einstein distribution
\begin{equation}\label{n1}
\bar{n}=n_B(\Omega,T)=\left[\exp\left(\frac{\Omega}{T}\right)-1\right]^{-1},
\end{equation}
and $\gamma$ is the decay rate, which is related to the spectral density of the bath $J(\omega)=\sum_jg_j^2\delta(\omega_j-\omega)$ via
\begin{equation}\label{gamma1}
\gamma=\pi J(\Omega).
\end{equation}
Note that the shift $\Delta$ is independent of the temperature, and although its effect is typical small (e.g. \cite{BreuerPetruccione,Carmichael}) we will not neglect it in our study. For an ohmic spectral density the frequency shift is
\[
\Delta=\alpha\mathrm{P.V.}\int^\infty_0 d\omega \frac{\omega e^{-\omega/\omega_c}}{\Omega-\omega}= \alpha\Omega  e^{-\Omega/\omega_c} \mathrm{Ei}\left(\Omega/\omega_c\right)-\alpha\omega_c,
\]
where $\mathrm{Ei}$ is the exponential integral function defined as
\[
\mathrm{Ei}(x)=-\mathrm{P.V.}\int^\infty_{-x}\frac{e^{-t}}{t}dt.
\]

In addition, note that the equation (\ref{Master1}) is Gaussian preseving \cite{Alessio}, as it is the limit of a linear interaction with an environment and so the total system remains Gaussian while the partial trace also preserves Gaussianity.

\subsection{Study of the Approximations}

\begin{figure}[!ht]
\centering
\includegraphics[width=0.75\textwidth]{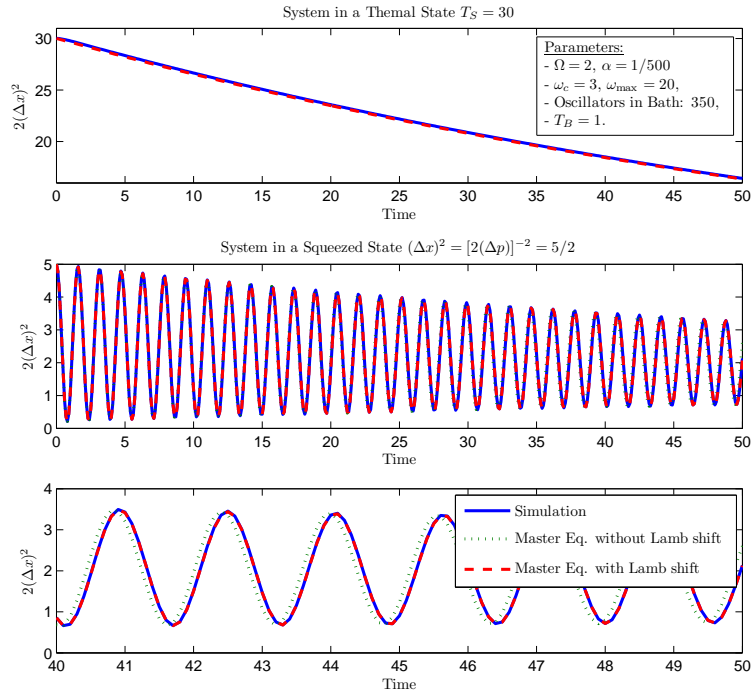}
\caption{Comparison of the evolution of $2(\Delta x)^2$ for an initially thermal and squeezed (vacuum) state. The bottom plot shows the effect of the Lamb shift, which produce a ``slippage'' in the squeezed state variances.}
\label{fig2}
\end{figure}

As a first step, we have plotted the variance of the $x$ coordinate for two different initial states of the system, these are a thermal and a squeezed state, see figure \ref{fig2}. The last plot clearly illustrates the closeness of the results for the Markovian master
equation, when compared to the effect of the Lamb shift. To explore this further, we now study several effects which pertain to the validity of this equation, by calculating the distance (in terms of the fidelity) between the simulated state $\rho_S^{(s)}$ and the state generated by the Markovian master equation $\rho_S^{(m)}$.

\subsubsection{Discreteness of the bath}

\begin{figure}[!ht]
\centering
\includegraphics[width=0.6\textwidth]{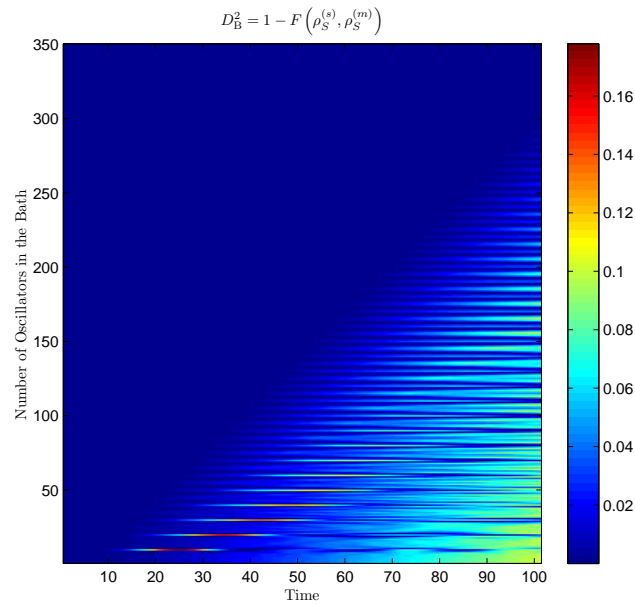}
\caption{Color map showing the dependency of the recurrence times with the size of the bath. The rest of the parameters are the same as in figure \ref{fig2}.}
\label{fig3}
\end{figure}

Due to the finite number of oscillators in the bath, we can only simulate inside a bounded time scale free of the back-action of the bath. This
produces revivals in the visualized dynamical quantities for times $t<\tau_R$, where $\tau_R$ is the recurrence time of the bath. Of course, the time after
which these revivals arise increases with the number of oscillators in the bath, and roughly speaking it scales as $\tau_R\propto M$. This behaviour is shown in figure \ref{fig3}, where the distance between the simulation and the Markovian master equation for a system initially in a thermal state with temperature $T_S=30$ is plotted as a function of the time and the number of oscillators.

\subsubsection{Temperature}

It is sometimes claimed that for ohmic spectral densities the Markovian master equation (\ref{Master1}) is not valid at low temperatures \cite{Carmichael,Weiss}. Of course, one must make clear the context in which this claim is made, and so for definiteness, let us focus on the validity with respect to the bath temperature. A detailed discussion of this situation can be found in the book by Carmichael \cite{Carmichael}. There the argument is based on the width of the correlation function $C_{12}(\tau)=\mathrm{Tr}[\tilde{B}_1 (s) B_2 \rho_\mathrm{th}]$, where $B_1^\dagger=B_2=\sum_{j=1}^Mg_ja_j$, which  increases for an Ohmic spectral density as the bath temperature decreases. More specifically, in the derivation of the Markovian master equation two kinds of correlation functions appear,
\begin{eqnarray*}
\fl C_{12}(s)=\mathrm{Tr}[\tilde{B}_1 (s) B_2\rho_\mathrm{th}]=\sum_{j,k}g_kg_je^{i\omega_js}\mathrm{Tr}[a^\dagger_j a_k\rho_\mathrm{th}]\\
=\sum_j^Mg_j^2e^{i\omega_js}\bar{n}(\omega_j,T),
\end{eqnarray*}
and
\begin{eqnarray*}
\fl C_{21}(s)=\mathrm{Tr}[\tilde{B}_2 (s) B_1 \rho_\mathrm{th}]=\sum_{j,k}g_kg_je^{-i\omega_js}\mathrm{Tr}[a_j a_k^\dagger \rho_\mathrm{th}]\\
=\sum_{j}^Mg_j^2e^{-i\omega_js}[\bar{n}(\omega_j,T)+1].
\end{eqnarray*}
We may call $C_{12}(s)\equiv C(-s,T)$ and $C_{21}(s)\equiv C(s,T)+C_0(s)$, and so in the continuous limit
\[
C_0(s)=\int_0^\infty J(\omega)e^{-i\omega s}d\omega=\alpha\int_0^\infty \omega e^{-i\omega(s-\omega_c^{-1})}d\omega=\frac{\alpha\omega_c^2}{(is \omega_c+1)^2},
\]
and
\[
C(s,T)=\int_0^\infty J(\omega)e^{-i\omega s}\bar{n}(\omega,T)d\omega=\alpha T^2\zeta\left(2,1-i s T+\frac{T}{\omega_c}\right),
\]
where here $\zeta(z,q)=\sum_{k=0}^\infty\frac{1}{[(q+k)^2]^{z/2}}$ is the so-called Hurwitz Zeta function, which is a generalization of the Riemann zeta function $\zeta(z)=\zeta(z,1)$ \cite{zeta}.

\begin{figure}[h]
\centering
\includegraphics[width=\textwidth]{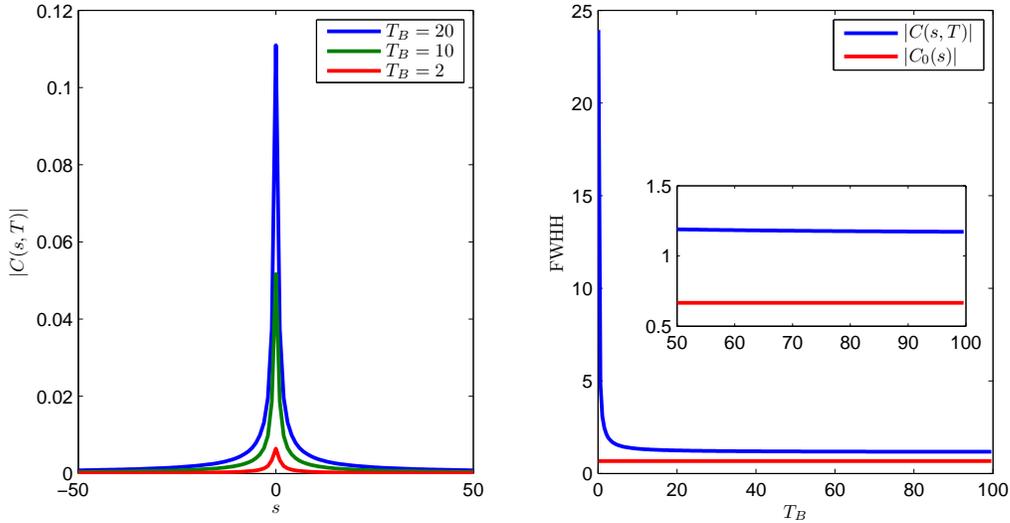}
\caption{On the left, the absolute value of the correlation function is plotted for several temperatures while the FWHH as a function of temperature is represented on the right.}
\label{fig4}
\end{figure}

In the left plot of figure \ref{fig4}, the absolute value of $C(s,T)$ is plotted for different temperatures. Note that the spreading of the correlation function is mainly caused by its ``height'' decrease, that is, in the limit $T\rightarrow0$, $C(s,T)\rightarrow0$. So one may also expect that the contribution of these correlations to the motion becomes less important as $T\rightarrow0$, in such a way that the problem of the infinite width can be counteracted, and this is indeed what seems to happen. To visualize this more carefully we have plotted in the right of figure \ref{fig4} the full weight at half height (FWHH) for both $C_0(s)$ and $C(s,T)$. In order to make valid the Markovian approximation, the typical time scale for the evolution of the system due to its interaction with the bath $\tau_S$ must be large in comparison with the decay time $\tau_B$ of the correlation functions. Loosely speaking, this can be characterized by the FWHH.

From figure \ref{fig4} one sees that for small temperatures $\tau_B$ (i.e. FWHH) is quite large, so it is expected that the Markovian approximation breaks down for values of $T$ such that $\tau_S\lesssim\tau_B$. However if $\alpha$ is small enough this will happen for values where the contribution of $C(s,T)$ to the convolution integrals is negligible in comparison with the contribution of $C_0(s)$, whose FWHH will remain constant and small with respect to $\tau_S$. As a rough estimation, using the parameters in figure \ref{fig2}, we find that to get a value of the FWHH comparable with $\tau_S\sim1/\sqrt{\alpha}\sim22.4$, we need a temperature of at least $T\sim0.05$. Both contributions enter in the Markovian master equation derivation via some convolution with the quantum state and one oscillating factor. We may get a very informal idea of how both contributions matter by looking at
their maximum values at $s=0$, for example $C(s=0,T=0.05)=3.27391\times10^{-7}$ and $C_0(s=0)=0.018$, and so it is clear that $C(s,T=0.05)$ will not have
a large effect on the dynamics. For large temperatures the FWHH of $C(s,T)$ remains small though now larger than $C_0(s)$, so it is expected that in the limit of high temperatures the accuracy of the Markovian master equation stabilizes to a value only a little worse than for $T=0$.

\begin{figure}[h]
\centering
\includegraphics[width=0.6\textwidth]{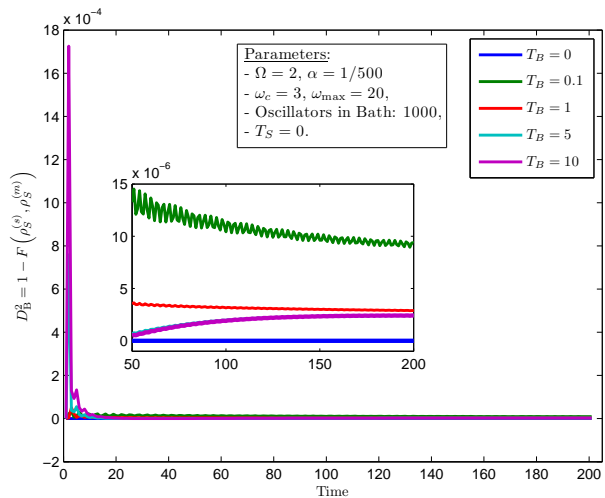}
\caption{Fidelity between the simulated state $\rho_S$ and that given by the Markovian master equation time evolution $\rho_M$, for several temperatures. For large times (see inset plot) temperature does not play a very significant role in the accuracy while for small times the accuracy depends mainly on the choice of the initial state of the system (see discussion in the text).}
\label{fig5}
\end{figure}

All of these conclusions are illustrated in figure \ref{fig5}, where the fidelity between the state from
the simulation and that from the Markovian master equation is plotted. The behaviour at very early times is mainly related to the choice of the initial state of the system, and reflects how it adjusts to the state of the bath under
the Markovian evolution \cite{Silbey}, different tendencies have been founded depending on the choice of initial state. However the behaviour with temperature is visible at longer times (since $\tau_B\sim\mathrm{FHWW}$ increases with $T$) which is in agreement with the conclusions drawn from the correlation functions (see small subplot). At zero temperature (blue line) the results are in closest
agreement, however, as the temperature is increased to $T=0.1$ the correlation
function broadens, which leads to a degradation (albeit small) in the modelling
precision. As the temperature increases further, the influence of this correlation function becomes more important and the FWHH decreases to a limiting value (see the plot on the right of figure \ref{fig4}), this convergence is reflected by the red, cyan and purple lines which show that the accuracy at large temperatures stabilizes to only a little worse than that at $T=0$, as was expected from figure \ref{fig4}.

In summary, the Markovian master equation (\ref{Master1}) does not properly describe the stimulated emission/absorption processes (the ones which depend on $C(s,T)$) for low temperatures, however the temperatures when this discrepancy is apparent are so small that the contribution from stimulated process are negligible in comparison with spontaneous emission, and so the discrepancy with the Markovian master equation is never large.

\subsubsection{Assumption of factorized dynamics $\rho(t)=\rho_S(t)\otimes\rho_\mathrm{th}$} \label{sectionproduct}

\begin{figure}[h]
\centering
\includegraphics[width=\textwidth]{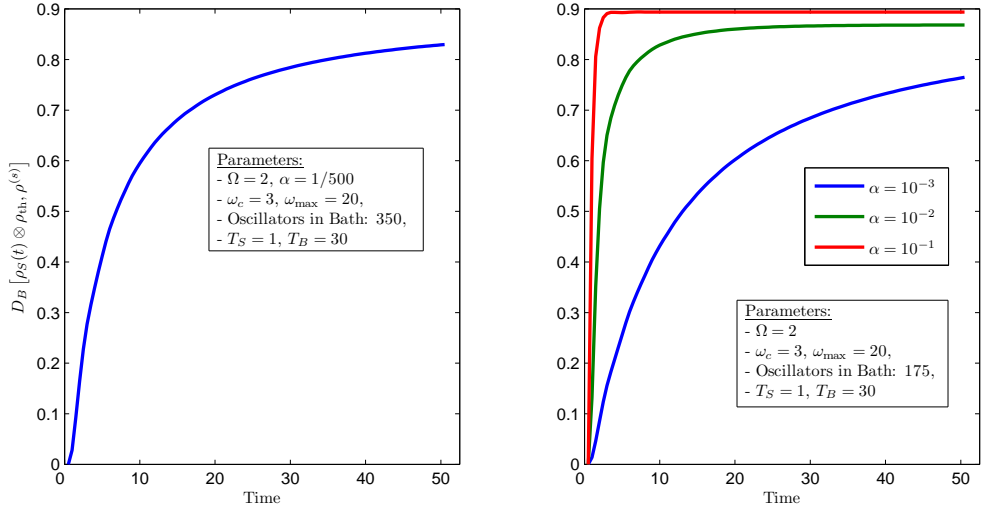}
\caption{Distance between the simulated states $\rho_S(t)\otimes\rho_{\mathrm{th}}$ and $\rho(t)$ as a function of time; on the left side, different numbers of oscillators in the bath are plotted (obtaining the same result on that time scale), and on the right side, different values of the coupling constant have been taken.}
\label{fig5.0}
\end{figure}

In the derivation of the Markovian master equation, one can arrive at equation (\ref{Bornapprox}) by iterating the Von-Neumann equation (\ref{Von-Neumann}) twice and assuming that the whole state factorizes as $\rho(t)\approx\rho_S(t)\otimes\rho_{\mathrm{th}}$ at any time (\cite{BreuerPetruccione,ModernCohen,Carmichael,Cohen}). This assumption has to be understood as an effective model for arriving at equation (\ref{Bornapprox}) without the use of projection operator techniques, however it does not make sense to assume that the physical state of the system is really a factorization for all time. Taking advantage of the ability to simulate the entire system we have plotted the distance between the simulated whole state $\rho(t)$ and the ansatz $\rho_S(t)\otimes\rho_{\mathrm{th}}$ as a function of time, see figure \ref{fig5.0}. On the left we have plotted the distance for $M=350$ oscillators in the bath, actually we have checked from several simulations that the results turn out to be independent of the number of oscillators as long as the maximum time is less than the recurrence time of the system. From figure \ref{fig3} we see that $t=50$ is less than the recurrence time for $M=175$, and so we have used this value and plotted the distance for different coupling strengths on the right.
It is clear that this distance is monotonically increasing in time (strictly,
in the limit of an environment with infinite degrees of freedom), and the
slope decreases with coupling strength. In section \ref{sectionmaster} we pointed out that the weak coupling approach make sense if the coupling is small and the environment has infinite degrees of freedom. This fits with the usual argument to take $\rho\approx\rho_S(t)\otimes\rho_{\mathrm{th}}$ in more informal derivation of Markovian master equations, that is ``the state of the environment is not so affected by the system'', but we stress again that this is an effective approach, without any physical meaning on the real state $\rho$.
%--------------------------------------------------------------------------------------------------------------------------------------%
%--------------------------------------------------------------------------------------------------------------------------------------%
%--------------------------------------------------------------------------------------------------------------------------------------%

\section{Two Coupled Damped Harmonic Oscillators}\label{section2HO}

\begin{figure}[h]
\centering
\includegraphics[width=0.7\textwidth]{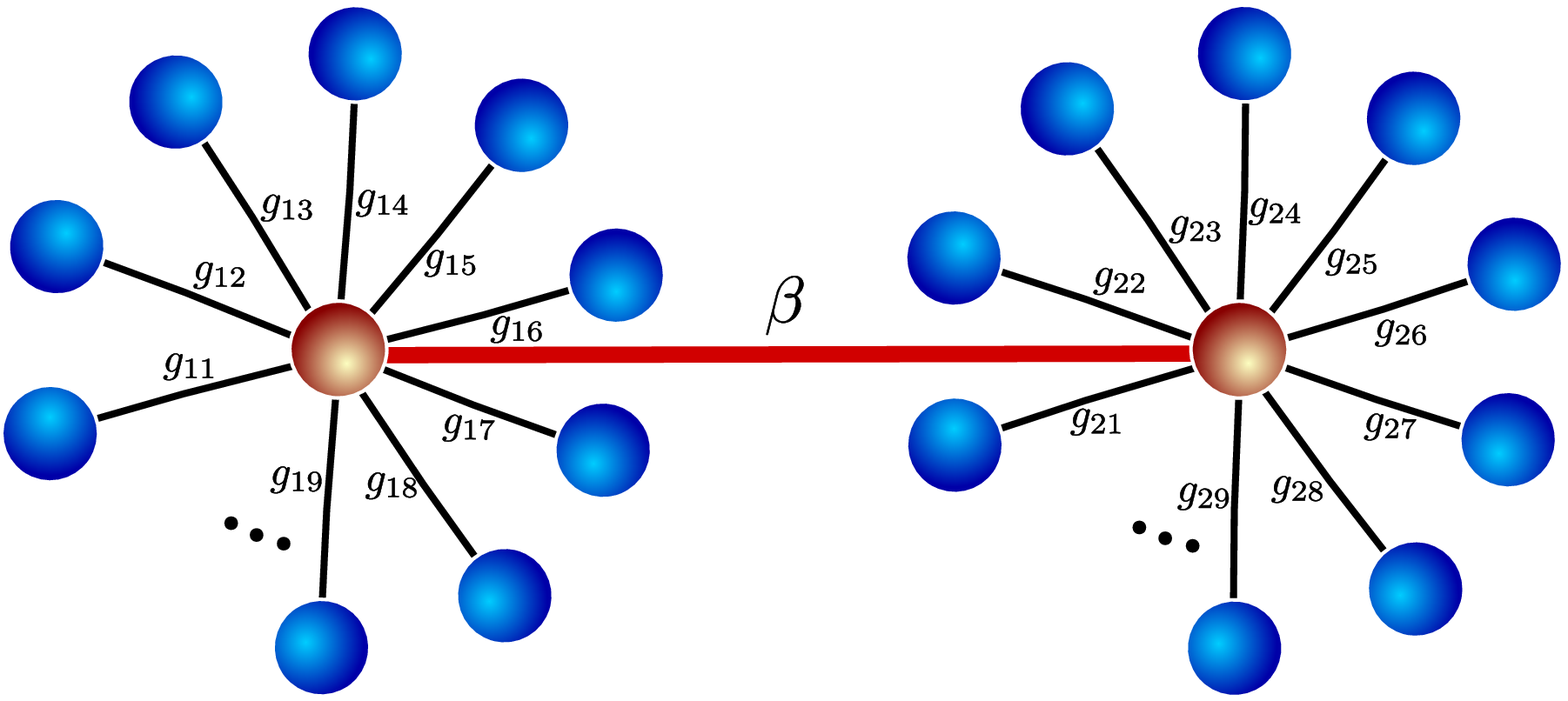}
\caption{The same model as figure \ref{fig0} for the case of two damped harmonic oscillators coupled together with strength $\beta$.}
\label{fig5.1}
\end{figure}

We now consider two coupled harmonic oscillators, which for simplicity we
take to have the same frequency $\Omega_1=\Omega_2=\Omega$, and each locally damped by their own reservoir (see figure \ref{fig5.1}), the Hamiltonian of the whole system is
\begin{equation}
H=H_{01}+H_{02}+V_{12}+H_{B1}+H_{B2}+V_{1B1}+V_{2B2},
\end{equation}
where the free Hamiltonians are given by
\begin{eqnarray}
H_{01}&=&\Omega a_1^\dagger a_1, \quad H_{02}=\Omega a_2^\dagger a_2,\nonumber\\
H_{B1}&=&\sum_{j=1}^M\omega_{1j}a_{1j}^\dagger a_{1j},\quad H_{B2}=\sum_{j=1}^M\omega_{2j}a_{2j}^\dagger a_{2j},\nonumber
\end{eqnarray}
with the couplings to the baths,
\begin{eqnarray}
V_{1B1}=\sum_{j=1}^Mg_{1j}(a_1^\dagger a_{1j} + a_1 a^\dagger_{1j}),\nonumber\\
V_{2B2}=\sum_{j=1}^Mg_{2j}(a_2^\dagger a_{2j} + a_2 a^\dagger_{2j}),\nonumber
\end{eqnarray}
and the coupling between oscillators,
\[
V_{12}=\beta(a_1^\dagger a_{2} + a_1 a^\dagger_{2}).
\]
Again we have employed the rotating wave approximation, and so we assume $\Omega\gg\beta$. For the case of $\Omega\sim\beta$ we must keep the antirotating terms $a_1a_2$ and $a_1^\dagger a_2^\dagger$. However note that the eigenfrequencies of the normal modes become imaginary if $\omega<2\beta$ (see for example \cite{Oscillators68}) and the system then becomes unstable, so even when keeping the antirotating terms, we must limit $\beta$ if we wish to keep the oscillatory behaviour.

\subsection{Exact Solution} \label{section2exact}
For the exact solution, the extension to two oscillators follows closely
that of a single damped harmonic oscillator. Again, we work in the Heisenberg
picture, and wish to solve for the vector $A=(a_1,a_{11},\ldots,a_{1M},a_2,a_{21},\ldots,a_{2M})^{\mathrm{T}}$,
given the differential equation,
\begin{equation}\label{odematrix2}
i\dot{A}=WA,
\end{equation}
where $W$ is now given by the matrix
\begin{equation}
W=\left(\begin{array}{cccccccc}
\Omega_1 & g_{11}      & \cdots & g_{1M}     & \beta    &            &        &           \\
g_{11}   & \omega_{12} &        &            &          &            &        &           \\
\vdots   &             &\ddots  &            &          &            &        &           \\
g_{1M}   &             &        &\omega_{1M} &          &            &        &           \\
\beta    &             &        &            & \Omega_2 &  g_{21}    & \cdots & g_{2M}    \\
         &             &        &            & g_{21}   &\omega_{21} &        &           \\
         &             &        &            & \vdots   &            &\ddots  &           \\
         &             &        &            & g_{2M}   &            &        &\omega_{2M}\\
\end{array}\right).
\end{equation}
The simulation process is then analogous to that of section \ref{section1exact}.

\subsection{Markovian Master Equations}

Unfortunately, the derivation of a Markovian master equation for coupled systems introduces
a number of additional complications. If the oscillators are uncoupled $\beta=0$, it is obvious that the Markovian master equation for their joint density matrix will be a sum of expressions like (\ref{Master1}),
\begin{equation}\label{MasterFree}
\frac{d}{dt}\rho_S(t)=-i[\bar{\Omega}a_1^\dagger a_1+\bar{\Omega}a_2^\dagger a_2,\rho_S(t)]+\mathcal{D}_1[\rho_S(t)]+\mathcal{D}_2[\rho_S(t)],
\end{equation}
where
\begin{eqnarray} \label{Dissipator}
\fl \mathcal{D}_j[\rho_S(t)]=\gamma_j(\bar{n}_j+1)\left(2a_j\rho_S(t) a_j^\dagger-a_j^\dagger a_j\rho_S(t)-\rho_S(t) a_j^\dagger a_j\right) \nonumber \\
+\gamma_j \bar{n}_j\left(2a_j^\dagger\rho_S(t) a_j - a_ja_j^\dagger \rho(t)-\rho_S(t) a_ja_j^\dagger \right),
\end{eqnarray}
here each frequency shift, decay rate and number of quanta are individually computed via equations (\ref{n1}), (\ref{gamma1}) and (\ref{shift1}) for each bath $j$. However for finite intercoupling we split the analysis in two subsections.

\subsubsection{Small intercoupling $\beta$} \label{sectionmasteraprox}

If $\beta$ is sufficiently small to not affect the shift and decay rates, one can expect a Markovian master equation of the form
\begin{equation}\label{MasterAprox}
\frac{d}{dt}{\rho}_S(t)=-i[\bar{\Omega}a_1^\dagger a_1+\bar{\Omega}a_2^\dagger a_2+V_{12},\rho_S(t)]+\mathcal{D}_1[\rho_S(t)]+\mathcal{D}_2[\rho_S(t)],
\end{equation}
an example of which for coupled subsystems can be found in \cite{ORHF09}, and we have given the details of a derivation based on projection operators in \ref{appendixSmallbeta}. In addition, this kind of approximation is often made in other contexts such as with damped systems driven by a classical field \cite{Carmichael}. Such a case will be analyzed in detail in section \ref{DrivenDampedHO}.

\subsubsection{Large intercoupling $\beta$}
To go further we must work in the interaction picture generated by the Hamiltonian $H_0=H_{\mathrm{free}}+V_{12}$ and apply the procedure described in section \ref{sectionmaster}. The details of the derivation are left for \ref{appendixLargebeta}, what is important however, is that the non-secular terms oscillate with a phase $e^{\pm 2i\beta t}$ so in order to neglect them we must impose $\beta\gg\alpha$, therefore the resultant equation is, in some sense, complementary to (\ref{MasterAprox}) valid if $\alpha\gtrsim\beta$. The final Markovian master equation in this regime takes the form

\begin{eqnarray}\label{MasterLbeta}
\fl \frac{d}{dt}\rho_S(t)=-i[\bar{\Omega}a_1^\dagger a_1+\bar{\Omega}a_2^\dagger a_2+\bar{\beta}\left(a_1a_2^\dagger+a_1^\dagger a_2\right),\rho_S(t)]\nonumber\\
+\sum_{j,k}^2K_{jk}^{(E)}
\left[a_j\rho_S(t)a_k^\dagger+\frac{1}{2}\{a_k^\dagger a_j,\rho_S(t)\}\right]\nonumber\\
+\sum_{j,k}^2K_{jk}^{(A)}\left[a^\dagger_j\rho_S(t)a_k+\frac{1}{2}\{a_k a_j^\dagger,\rho_S(t)\}\right],
\end{eqnarray}
here
\begin{eqnarray*}
\bar{\Omega}&=&\Omega+[\Delta_1(\Omega_+)+\Delta_2(\Omega_+)+\Delta_1(\Omega_-)+\Delta_2(\Omega_-)]/4,\\
\bar{\beta}&=&\beta+[\Delta_1(\Omega_+)+\Delta_2(\Omega_+)-\Delta_1(\Omega_-)-\Delta_2(\Omega_-)]/4,
\end{eqnarray*}
and $K_{jk}^{(E)}$ and $K_{jk}^{(A)}$ are two positive semidefinite Hermitian matrices with coefficients
\begin{eqnarray}
\fl K_{11}^{(E)}=K_{22}^{(E)}=\{\gamma_1(\Omega_+)[\bar{n}_1(\Omega_+)+1]+\gamma_2(\Omega_+)[\bar{n}_2(\Omega_+)+1]\nonumber \\
+\gamma_1(\Omega_-)[\bar{n}_1(\Omega_-)+1]+\gamma_2(\Omega_-)[\bar{n}_2(\Omega_-)+1]\}/2,\\
\fl K_{12}^{(E)}=K_{21}^{(E)\ast}=\{\gamma_1(\Omega_+)[\bar{n}_1(\Omega_+)+1]+\gamma_2(\Omega_+)[\bar{n}_2(\Omega_+)+1]\nonumber\\
-\gamma_1(\Omega_-)[\bar{n}_1(\Omega_-)+1]-\gamma_2(\Omega_-)[\bar{n}_2(\Omega_-)+1]\}/2,
\end{eqnarray}
\begin{eqnarray}
\fl K_{11}^{(A)}=K_{22}^{(A)}=[\gamma_1(\Omega_+)\bar{n}_1(\Omega_+)+\gamma_2(\Omega_+)\bar{n}_2(\Omega_+)\nonumber \\
+\gamma_1(\Omega_-)\bar{n}_1(\Omega_-)+\gamma_2(\Omega_-)\bar{n}_2(\Omega_-)]/2,\\
\fl K_{12}^{(A)}=K_{21}^{(A)\ast}=[\gamma_1(\Omega_+)\bar{n}_1(\Omega_+)+\gamma_2(\Omega_+)\bar{n}_2(\Omega_+)\nonumber\\
-\gamma_1(\Omega_-)\bar{n}_1(\Omega_-)-\gamma_2(\Omega_-)\bar{n}_2(\Omega_-)]/2,
\end{eqnarray}
where $\gamma_j$, $\Delta_j$ and $\bar{n}_j$ are evaluated according to the spectral density and temperature of the bath $j$ and $\Omega_\pm=\Omega\pm\beta$.

\subsection{Study of the Approximations}
\begin{figure}[h]
\centering
\includegraphics[width=\textwidth]{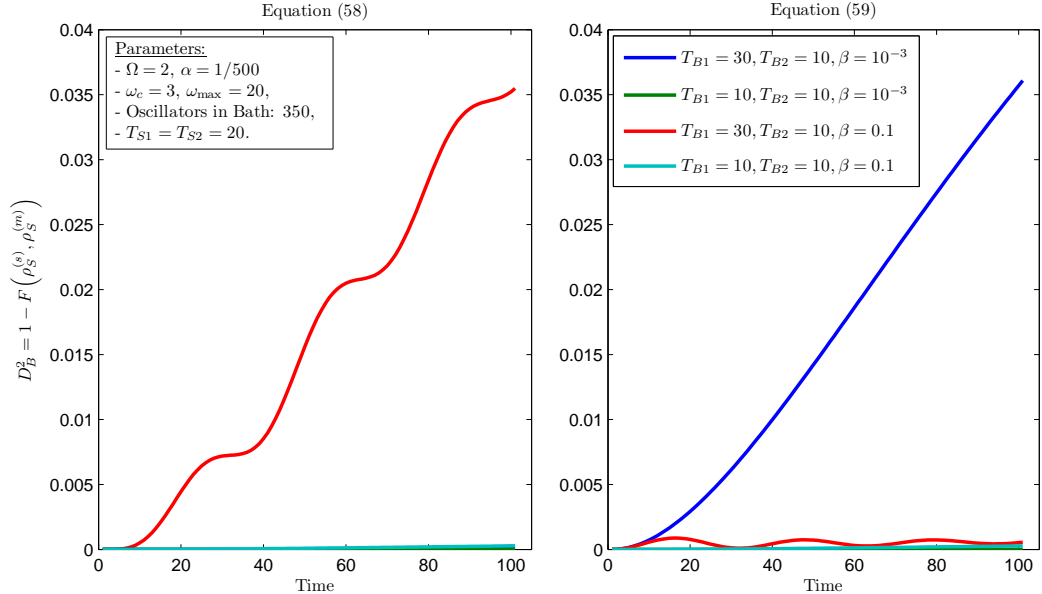}
\caption{On the left, the fidelity between the simulated state $\rho_S^{(s)}$ and that according to the Markovian master equation (\ref{MasterAprox}). The analog using the Markovian master equation (\ref{MasterLbeta}) is plotted on the right. In both plots
the parameters and legends are the same.}
\label{fig6}
\end{figure}

By virtue of the derivation, equations (\ref{MasterAprox}) and (\ref{MasterLbeta}) preserve both complete positivity and Gaussianity (because they arise from a linear interaction with the environment). Thus  we can test their regimes of validity using simulations of Gaussian states, and the appropriate fidelity formulas. In figure \ref{fig6} we have plotted the fidelity between both states for the Markovian master equation (\ref{MasterAprox}) (left side) and for (\ref{MasterLbeta}) (right side).

\begin{figure}[h]
\centering
\includegraphics[width=0.65\textwidth]{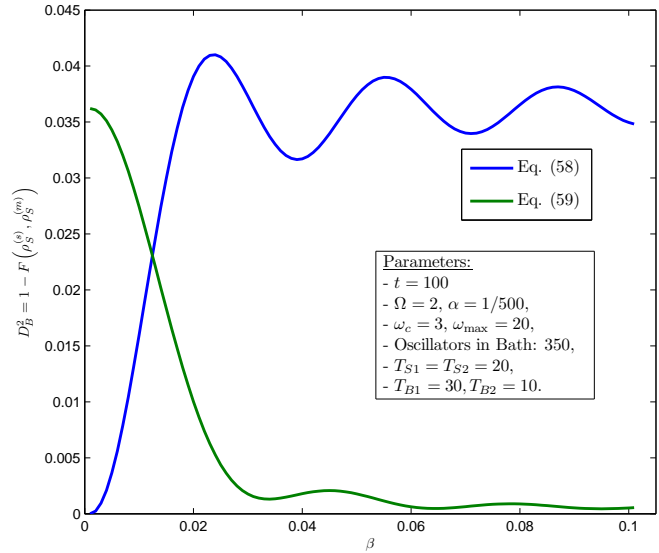}
\caption{Fidelity between the simulated state $\rho_S^{(s)}$ and $\rho_S^{(m)}$ according to the Markovian master equations (\ref{MasterAprox}) and (\ref{MasterLbeta}) at fixed time as a function of the coupling between the damped oscillators.}
\label{fig7}
\end{figure}

From these results one concludes that when modeling a system with multiple
baths at different temperatures equations (\ref{MasterAprox}) and (\ref{MasterLbeta})
are each accurate in their theoretically applicable regimes. However, for
baths at the same temperature, it seems both equations give good results.
A natural, and important, question is to ask is whether an intermediate range
of couplings exist, such that neither (\ref{MasterAprox}) or (\ref{MasterLbeta})
give useful results. In figure \ref{fig7} the fidelity between the simulation and the Markovian master equation states have been plotted for both equations at fixed time $t=100$ as a function of the intercoupling strength $\beta$.

We see that for the parameters shown on the plot, there is a small range between $\beta\sim0.01-0.02$ where neither Markovian master equation obtains a high
precision. However, note that this range becomes smaller as the coupling with the bath decreases, and so generally both master equations cover a good range of values of $\beta$.

\subsubsection{Baths with the same temperature}

We now examine the role of the bath temperatures in more detail. Since the simulations seem to produce good results for both Markovian master equations when the temperature of the local baths are the same, regardless of the strength of the intercoupling, it is worth looking at why this happens. In the case of
equation (\ref{MasterLbeta}) it is reasonable to expect that this will remain
valid for small $\beta$, because when $\beta\rightarrow0$ this equation approaches (\ref{MasterAprox}) if the bath temperatures and spectral densities are the same. That is, the off-diagonal terms of the matrices $K^{(E)}$ and $K^{(A)}$ do not contribute much, $\bar{\beta}\sim\beta$ and the rest of coefficients become approximately equal to those in (\ref{MasterAprox}.) Note this only happens under these conditions.

\begin{figure}[h]
\centering
\includegraphics[width=0.7\textwidth]{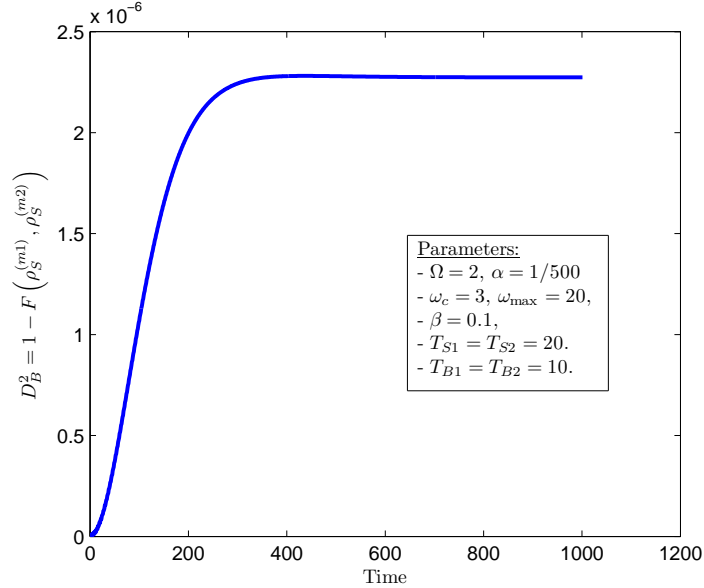}
\caption{Fidelity between states $\rho_S^{(m1)}$ and $\rho_S^{(m2)}$ corresponding to Markovian master equations (\ref{MasterAprox}) and (\ref{MasterLbeta}) respectively.}
\label{fig8}
\end{figure}

Essentially the same argument applies to equation (\ref{MasterAprox}) in
the large $\beta$ limit. On the one hand, for a relatively small value of $\beta$ ($=0.1$) in comparison to $\omega$, the off-diagonal elements of the matrices $K^{(E)}$ and $K^{(A)}$ in the master equation (\ref{MasterLbeta})
are unimportant in comparison with the diagonals. On the other hand, the diagonal terms are also alike for the same reason, and so both master equations will be quite similar. However note that at later times the behaviour of both equations start to differ, and the steady states are not the same. By
construction, the steady state of equation (\ref{MasterLbeta}) is the thermal
state of the composed system \cite{BreuerPetruccione,Davies1}, whereas that
of master equation (\ref{MasterAprox}) is not (although it tends to the thermal state as $\beta\rightarrow0$ of course). Surprisingly the divergences between both equations, even for large times, are actually very small, see figure \ref{fig8}. In some cases, while the steady state of (\ref{MasterAprox})
is not strictly thermal, the fidelity with that of (\ref{MasterLbeta}) is
 more than 99.999\%.

%--------------------------------------------------------------------------------------------------------------------------------------%
%--------------------------------------------------------------------------------------------------------------------------------------%
%--------------------------------------------------------------------------------------------------------------------------------------%

\section{Driven Damped Harmonic Oscillator}\label{DrivenDampedHO}

\begin{figure}[h]
\centering
\includegraphics[width=0.4\textwidth]{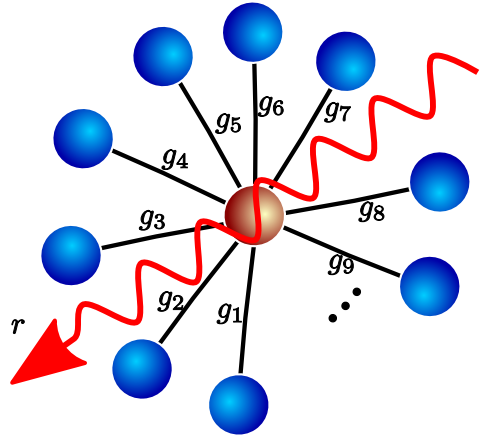}
\caption{A single damped oscillator interacting with a classical incident field with Rabi frequency $r$.}
\label{fig8.1}
\end{figure}

One situation which is also interesting to analyze is that of adding a driving
term in the Hamiltonian of the damped oscillator. At this stage we consider again one single oscillator, damped by a thermal bath and driven by a coherent field (figure \ref{fig8.1}). This is described by a semiclassical Hamiltonian in the rotating wave approximation:
\begin{equation}
H(t)=\Omega a^\dagger a+r(a^\dagger e^{-i\omega_Lt}+a e^{i\omega_Lt})+ \sum_{j=1}^M\omega_ja^\dagger_j a_j+\sum_{j=1}^Mg_j(a^\dagger a_j + a a^\dagger_j),
\end{equation}
here $\omega_L$ is the frequency of the incident field and $r$ the Rabi frequency.

\subsection{Exact Solution} \label{section3exact}
To obtain the exact solution of this system let us consider for a moment the Schr\"odinger picture,
\[
\frac{d|\psi(t)\rangle}{dt}=-iH(t)|\psi(t)\rangle.
\]
We solve this equation by means of the unitary transformation $U_{\mathrm{rot}}(t)=e^{iH_{\mathrm{rot}}t}$ where $H_{\mathrm{rot}}=\omega_L \left(a^\dagger a+\sum_{j=1}^M a^\dagger_j a_j\right)$. Making the substitution $|\tilde{\psi}(t)\rangle=U_{\mathrm{rot}}(t)|\psi(t)\rangle$ we immediately obtain
\[
\frac{d|\tilde{\psi}(t)\rangle}{dt}=i[H_{\mathrm{rot}}-U_{\mathrm{rot}}(t)H(t)U^\dagger_{\mathrm{rot}}(t)]|\tilde{\psi}(t)\rangle=-iH_0|\tilde{\psi}(t)\rangle,
\]
where $H_0=(\Omega-\omega_L)a^\dagger a+r(a+a^\dagger)+\sum_{j=1}^M(\omega_j-\omega_L)a^\dagger_j a_j+\sum_{j=1}^Mg_j(a^\dagger a_j + a a^\dagger_j)$ is time-independent. Returning to the Schr\"odinger picture, the evolution of the states is then,
\[
|\psi(t)\rangle=U(t,0)|\psi(0)\rangle=e^{-iH_{\mathrm{rot}}t} e^{-iH_0t} |\psi(0)\rangle.
\]
In order to avoid differential equations with time-dependent coefficients, we can study the evolution in a X-P time rotating frame; in that frame the annihilation (and creation) operators $\tilde{a}=e^{-iH_{\mathrm{rot}}t}ae^{iH_{\mathrm{rot}}t}$ will evolve according to
\[
\tilde{a}(t)=U^\dagger(t,0)e^{-iH_{\mathrm{rot}}t}ae^{iH_{\mathrm{rot}}t}U(t,0)=e^{iH_0t}ae^{-iH_0t}.
\]
That is
\begin{eqnarray}\label{annihirotante}
i\dot{\tilde{a}}=[\tilde{a},H_0]=(\Omega-\omega_L) \tilde{a}+\sum_{j=1}^Mg_j \tilde{a}_j+r,\\
i\dot{\tilde{a}}_j=[\tilde{a}_j,H_0]=(\omega_j-\omega_L)\tilde{a}_j+g_j \tilde{a},
\end{eqnarray}
which is quite similar to (\ref{annihi}) but with the additional time-independent term $r$. Following the notation of section \ref{section1exact} we can write
\[
i\dot{\tilde{A}}=W_0\tilde{A}+b,
\]
here $b=(r,0,\ldots,0)^{\mathrm{T}}$ and $W_0$ is found from (\ref{Wmatrix})
as $W-\omega_L \mathds{1}$. The solution of this system of differential equations is
\[
\tilde{A}(t)=e^{-iW_0t}\left[A(0)-i\int_0^tdse^{iW_0s}b\right].
\]
If $W_0$ is invertible this equation can be written as
\begin{equation}
\tilde{A}(t)=e^{-iW_0t}\left[A(0)+W_0^{-1}b\right]-W_0^{-1}b,
\end{equation}
%and so for creation operators
%\begin{equation}
%\tilde{A}^\dagger(t)=e^{iW_0t}\left[A^\dagger(0)+W_0^{-1}b\right]-W_0^{-1}b.
%\end{equation}
Analogously to (\ref{X(t)}) and (\ref{Y(t)}) we find
\begin{eqnarray}
\tilde{X}(t)=T_R^0X(0)-T_I^0P(0)+T_R^0W_0^{-1}b-W_0^{-1}b,\\
\tilde{P}(t)=T_I^0X(0)+T_R^0P(0)+T_I^0W_0^{-1}b,
\end{eqnarray}
where $T_R^0$ and $T_I^0$ are as in (\ref{TRTI}) for $W_0$. Thus, by writing
\[
\mathcal{M}^0=\left(\begin{array}{cc}
T_R^0 & -T_I^0\\
T_I^0 & T_R^0
\end{array}\right), \quad \mathcal{B}=\left(\begin{array}{c}
(T_R^0-\id)W_0^{-1}b\\
T_I^0W_0^{-1}b
\end{array}\right),
\]
we find that the position and momentum expectation values evolve as
\begin{equation}
\tilde{R}(t)=\mathcal{M}^0R(0)+\mathcal{B}.
\end{equation}
Note that in this case the first moments of the state change, despite $\langle R(0)\rangle=0$. To calculate the evolution of the covariance matrix, we proceed in the same way as before,
\begin{eqnarray}
\fl \langle \tilde{R}_i(t) \tilde{R}_j(t)\rangle=\sum_{k,\ell}\mathcal{M}^0_{i,k}\mathcal{M}^0_{j,\ell}\langle R_i(0)R_j(0)\rangle
+\sum_k\mathcal{M}^0_{i,k}\langle R_k(0)\rangle\mathcal{B}_j\nonumber \\
+\mathcal{B}_j\sum_k\mathcal{M}^0_{j,\ell}\langle R_\ell(0)\rangle+\mathcal{B}_i\mathcal{B}_j,
\end{eqnarray}
and analogously for the solutions for $\langle \tilde{R}_j(t) \tilde{R}_i(t)\rangle$ and $\langle \tilde{R}_i(t)\rangle \langle\tilde{R}_j(t)\rangle$. Combining
these terms, we find the $\mathcal{B}$ cancel and so, in a similar fashion to (\ref{C11}),(\ref{C1M+2}) and (\ref{CM+2M+2}),
\begin{eqnarray}
\tilde{\mathcal{C}}_{1,1}(t)=(\mathcal{M}^0_1,\mathcal{C}(0)\mathcal{M}^0_1),\nonumber \\
\tilde{\mathcal{C}}_{1,M+2}(t)=\tilde{\mathcal{C}}_{M+2,1}(t)=(\mathcal{M}^0_1,\mathcal{C}\mathcal{M}^0_{M+2})\nonumber\\
\tilde{\mathcal{C}}_{M+2,M+2}(t)=(\mathcal{M}^0_{M+2},\mathcal{C}\mathcal{M}^0_{M+2}),
\end{eqnarray}
where, of course, $\mathcal{M}^0_1$ and $\mathcal{M}^0_2$ are as in (\ref{M1M2}) for $\mathcal{M}^0$.

\subsection{Markovian Master Equations}

In order to derive a Markovian master equation for this system we must take account
of two important details. First, since the Hamiltonian is time-dependent the generator of the master equation must also be time-dependent,
\[
\frac{d\rho_S(t)}{dt}=\mathcal{L}_t\rho_S(t),
\]
whose solution defines a family of propagators $\mathcal{E}(t_2,t_1)$ such that
\begin{eqnarray*}
\rho_S(t_2)=\mathcal{E}(t_2,t_1)\rho_S(t_1),\\
\mathcal{E}(t_3,t_1)=\mathcal{E}(t_3,t_2)\mathcal{E}(t_2,t_1).
\end{eqnarray*}
These can be written formally as a time-ordered series
\[
\mathcal{E}(t_1,t_0)=\mathcal{T}e^{\int_{t_0}^{t_1}\mathcal{L}_{t'}dt'},
\]
where $\mathcal{T}$ is the well-known time-ordering operator. Similarly to the case of time-independent equations it can be shown that the family $\mathcal{E}(t_2,t_1)$ is completely positive for all $(t_2\geq t_1)$ if and only if $\mathcal{L}_t$ has the Kossakowski-Lindblad form for any time $t$ \cite{rivas}.

The second problem is that there is an absence of rigorous methods to arrive at a completely positive master equation in the Markovian limit when the Hamiltonian is time-dependent, with the exception of adiabatic regimes of external perturbations \cite{Ht}. Fortunately in this case, due to the simple periodic time-dependence of the Hamiltonian, we will be able to obtain Markovian master equations valid for large (to some degree) Rabi frequencies, even though the complexity of the problem has increased. In our derivation, we will distinguish between three cases: these will be when the Rabi frequency is very small; when the driving is far off resonance $(|\omega_L-\Omega|\gg0)$ and finally the identical
case without the secular approximation.

The details of the derivation are left for the \ref{appendixDriven}, but in these three cases we find a Markovian master equation with the structure
\[
\frac{d}{dt}{\rho}_S=-i[\bar{\Omega}a^\dagger a+\bar{r}e^{i\omega_Lt}a+\bar{r}^{\ast}e^{-i\omega_Lt}a^\dagger,\rho_S]+\mathcal{D}(\rho_S),
\]
where $\mathcal{D}$ is given by (\ref{Dissipator}), $\bar{\Omega}=\Omega+\Delta$ is the same as for a single damped oscillator, and $\bar{r}$ is a renormalized Rabi frequency due to the effect of the bath. Note that as the incident field
alters the position operator of the oscillator, which in turn couples to
the bath, one should expect that the field is itself also effected by the
environment. For small Rabi frequencies an argument similar to section \ref{sectionmasteraprox} gives simply
\begin{equation}\label{rverysmall}
\bar{r}=r,
\end{equation}
whereas, when the driving field is far from resonance, $|\omega_L-\Omega|\gg0$, we obtain
\begin{equation}\label{off-resonant}
\bar{r}=r\left[1+\frac{\Delta(\Omega)+i\gamma(\Omega)}{\Omega-\omega_L}\right].
\end{equation}
Finally, if we neglect the secular approximation, this regime yields
\begin{equation}\label{no-secular}
\bar{r}=r\left[1+\frac{\Delta(\Omega)+i\gamma(\Omega)}{\Omega-\omega_L}-\frac{\Delta(\omega_L)+i\gamma(\omega_L)}{\Omega-\omega_L}\right].
\end{equation}

Without entering into the details of the derivation, one sees that equations (\ref{off-resonant}) and (\ref{no-secular}) are problematic on resonance $|\Omega-\omega_L|\sim0$. This is due to two approximations, one is the secular approximation in (\ref{off-resonant}), and the other is the second order in the perturbative series. In the derivation in \ref{appendixDriven} it is clear why in this case the series diverges for $|\Omega-\omega_L|\sim0$.

\subsection{Study of the Approximations}\label{DrivenDampedHOSimulation}
%Note that it is not clear how one should expect the range of validity of
%each equation to vary. This is because we cannot now unambiguously choose a set of parameters in such a way that one equation.... [not sure about this]....

%One could suppose that the more elaborate equations (\ref{off-resonant})
%and (\ref{no-secular}) would provide the better approximation. However, there
%is still the question of how effective they are, and whether the additional
%effort required to obtain them is worthwhile in comparison to the simpler
%equation (\ref{rverysmall}).

Note that in this case the range of validity of each equation is now more ambiguous than in previous sections where we have dealt with undriven systems.
%because we cannot clearly chose a set of parameter in such a way that one equation is not valid and the rest are (the two only possibilities would be: to take $r\gg1$ holding the conditions for the equations (\ref{off-resonant}) and (\ref{no-secular}), and expect that (\ref{rverysmall}) fails; and the second possibility is in resonance $|w^L-\Omega|=0$ and small $r$ expecting that (\ref{rverysmall}) works),
Which one is more appropriate is going to be discovered by simulation, although one could suppose that the more elaborate equations (\ref{off-resonant}) and (\ref{no-secular}) would provide the better approximation. However, there is still the question of how effective they are, and whether the additional effort required to obtain them is worthwhile in comparison to the simpler equation (\ref{rverysmall}).

In addition note that in every case the covariance matrix is unaffected by the driving term, which only produce a change in the first moments. Furthermore,
as the fidelity is invariant under unitary operations, we are always free
to work in the frame rotating with the field. Therefore, all calculations
can be performed with the rotating observables.

\begin{figure}[h]
\centering
\includegraphics[width=\textwidth]{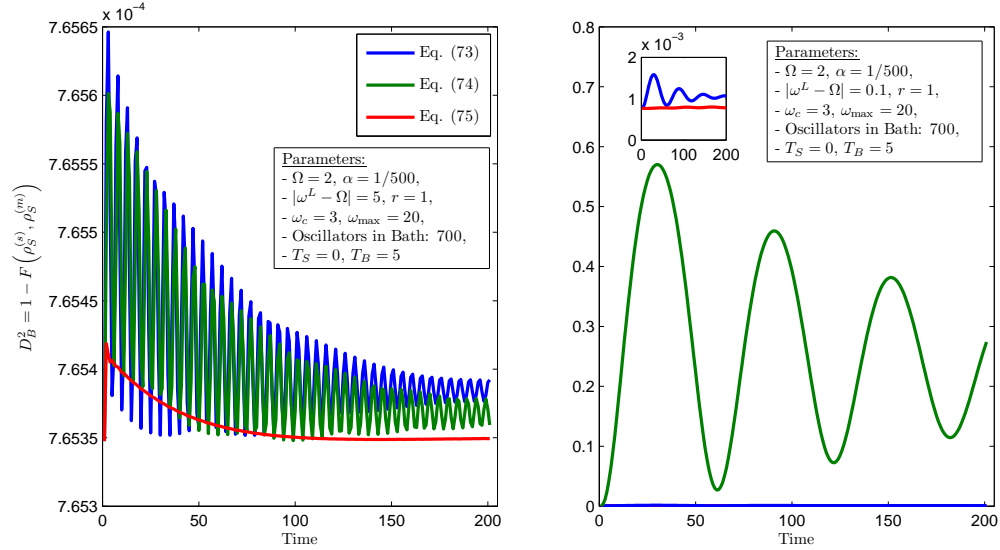}
\caption{Fidelity between $\rho_S^{(s)}$ and $\rho_S^{(m)}$ for different renormalized Rabi frequencies (\ref{rverysmall}), (\ref{off-resonant}) and (\ref{no-secular}). An example of off resonance is shown on the left, whereas
the plot on the right is close to resonance.}
\label{fig9}
\end{figure}

In figure \ref{fig9} the fidelities are plotted for close to and far from
resonance. Compare the amount of disagreement with the fidelity of a single damped oscillator in figure \ref{fig5}. For global features, the more elaborate
equation (\ref{no-secular}) works better in both cases, although the difference with (\ref{rverysmall}) is very small. As expected, the choice of (\ref{off-resonant}) is preferable to the choice of (\ref{rverysmall}) when out of resonance, but gives quite poor results when close to resonance. However, when off resonance the difference among the three choices is essentially small.

\begin{figure}[h]
\centering
\includegraphics[width=0.7\textwidth]{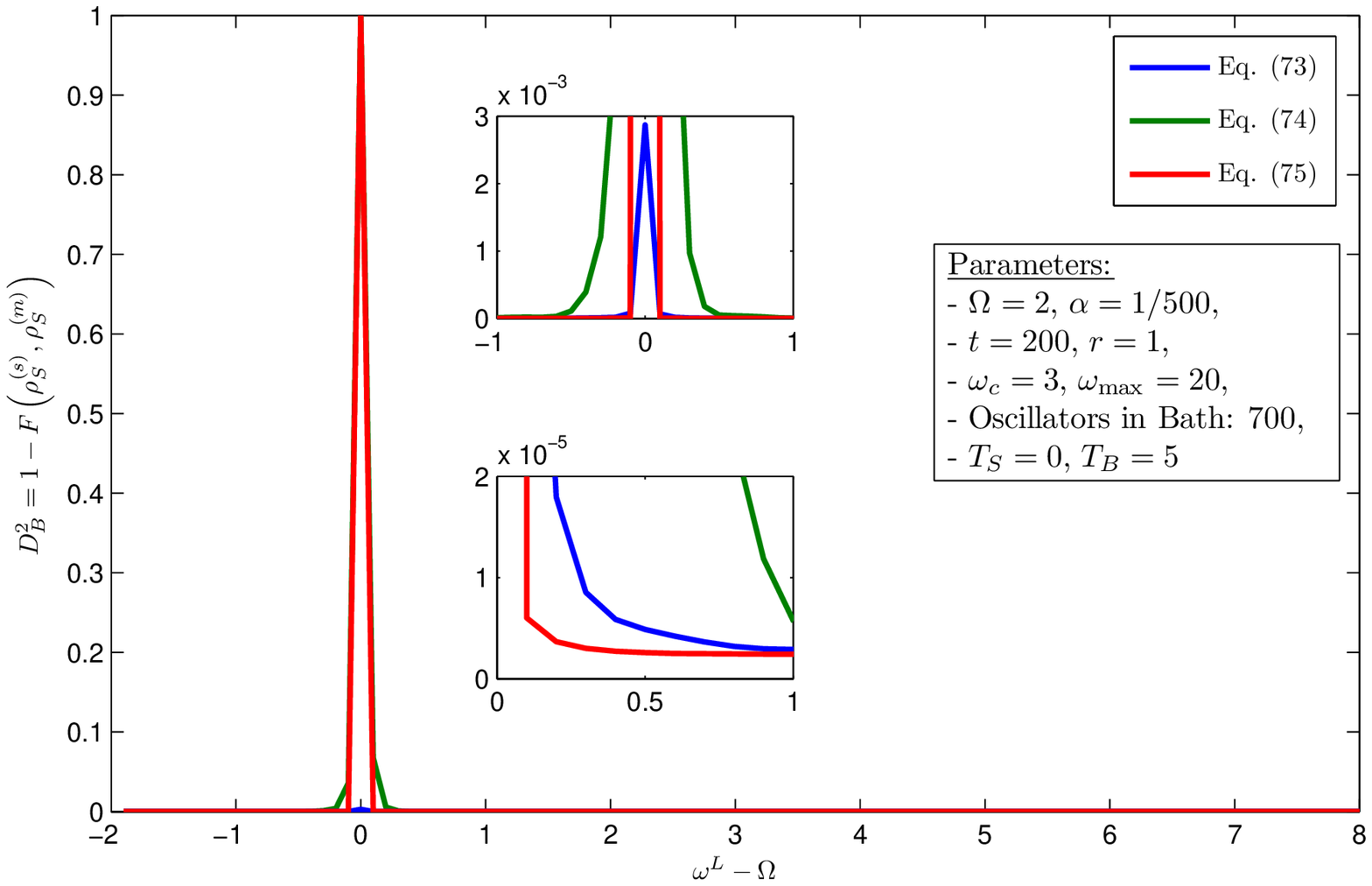}
\caption{Fidelity between $\rho_S^{(s)}$ and $\rho_S^{(m)}$ for different renormalized Rabi frequencies (\ref{rverysmall}), (\ref{off-resonant}) and (\ref{no-secular}) as a function of the detunning.}
\label{fig10}
\end{figure}

Given these results, it is worthwhile to look at how the fidelities at one fixed time vary as a function of the detunning, this is done in figure \ref{fig10} (note we choose a large value for the time, so we avoid the potentially confusing effect due to the oscillatory behaviour depicted in figure \ref{fig9}).

Here we see that both (\ref{off-resonant}) and (\ref{no-secular}) fail close
to resonance, as was expected from the perturbative approach. Equation (\ref{rverysmall}) gives good results due to the small Rabi frequency, however note in comparison
to (\ref{no-secular}) the accuracy quickly drops off as we move away from
$\omega_L - \Omega=0$. A similar effect can be seen when compared to (\ref{off-resonant})
for larger detunnings.

\begin{figure}[h]
\centering
\includegraphics[width=0.7\textwidth]{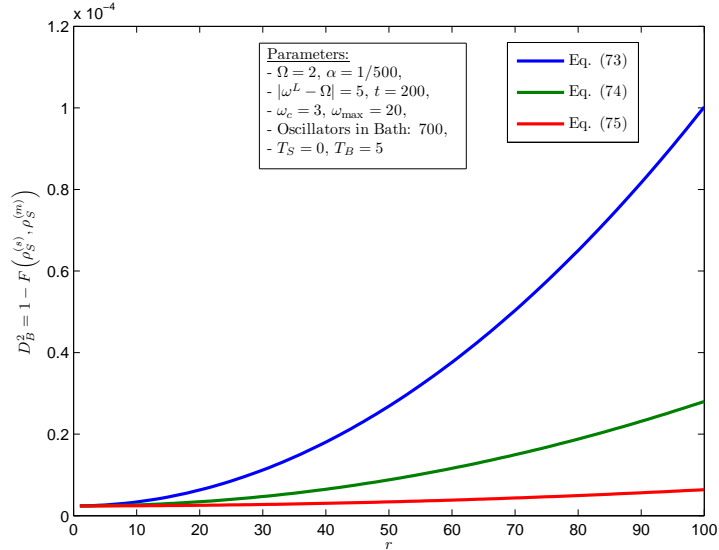}
\caption{Fidelity between $\rho_S^{(s)}$ and $\rho_S^{(m)}$ for different renormalized Rabi frequencies (\ref{rverysmall}), (\ref{off-resonant}) and (\ref{no-secular}) as a function of the Rabi frequency.}
\label{fig11}
\end{figure}

Finally, in figure \ref{fig11} we test the dependency of the fidelities on
the strength of the Rabi frequencies far from resonance. Here the worst behaviour
is observed for (\ref{rverysmall}), as expected.

In summary, for the case of a driven damped harmonic oscillator the difference in accuracy among Markovian master equations is generally small. Equations (\ref{off-resonant}) and (\ref{no-secular}) work better except in the case of resonance, where (\ref{rverysmall}) gives more accurate results, as long as the Rabi frequency is small. The justification to use one equation over another will depend on the context and the accuracy which one wants to obtain, but given that the differences are so small the simplest choice (\ref{rverysmall}) seems to be the more ``economical'' way to describe the dynamics.
%--------------------------------------------------------------------------------------------------------------------------------------%
%--------------------------------------------------------------------------------------------------------------------------------------%
%--------------------------------------------------------------------------------------------------------------------------------------%

\section{Conclusions}
We have obtained and studied the range of validity of different Markovian master equations for harmonic oscillators by means of exactly simulating the dynamics, and comparing the predictions with those obtained from evolving the system using the master equations. In particular,

\begin{itemize}
\item We have clarified the possible detrimental effect of low temperatures on the Markovian treatment of a damped oscillator, showing that the Markovian master equation provides good accuracy regardless of the temperature of the bath.
\item We have shown that the system-environment state factorization assumption for all times has to be understood in general as an effective model by deriving the same equation using the projection operator technique.
%And that the real whole state system plus environment is far from be factorized.
\item We analysed two strategies for finding completely positive Markovian master equations for two harmonic oscillators coupled together under the effect of local baths, indicating that both are complementary in their range of validity. Moreover, when the temperature of the local baths is the same the difference between them is quite small.
\item In the same spirit, we derived time inhomogeneous completely positive Markovian master equations for a damped oscillator which is driven by an external semi-classical field. We studied the validity of each one and pointed out that completely positive dynamics can be obtained even without secular approximation (for these kinds of inhomogeneous equations).
\end{itemize}
%On the other hand, in this work we have found no surprising deviations from what is expected based on theoretical arguments.
Despite the fact that we have focused on harmonic oscillator systems, the proposed method is general and we expect that non-harmonic systems should behave in a similar manner with respect to the validity of the equations. This suggest that the general conclusions made here are widely applicable to any other settings involving a weak interaction with an environment.

In this regard, we hope that the present study may help in providing a better understanding and a transparent description of noise in interacting systems, including those situations where the strength of the internal system interaction is large. There are currently many quantum scenarios open to the use of these techniques, including realizations of harmonic and spin chains in systems of trapped ions \cite{iones}, superconducting qubits \cite{scq} and nitrogen-vacancy (NV) defects in diamond \cite{nv}.

Moreover, interacting systems subject to local reservoirs have been recently
treated under the assumption of weak internal system interaction in theoretical studies
ranging from the excitation transport properties of biomolecules \cite{bio} to the
stability of topological codes for quantum information \cite{Topological}.
%--------------------------------------------------------------------------------------------------------------------------------------%
%--------------------------------------------------------------------------------------------------------------------------------------%
%--------------------------------------------------------------------------------------------------------------------------------------%

\ack
A.R. acknowledges Alex Chin for fruitful discussions. This work was supported by the STREP projects CORNER and HIP, the Integrated projects on QAP and Q-ESSENCE, the EPSRC QIP-IRC GR/S82176/0 and an Alexander von Humboldt Professorship.
%--------------------------------------------------------------------------------------------------------------------------------------%
%--------------------------------------------------------------------------------------------------------------------------------------%
%--------------------------------------------------------------------------------------------------------------------------------------%

\appendix

\section{Details of the simulation} \label{appendixsimulation}

In order to make an appropriate comparison between the exact evolutions, such as those in sections \ref{section1exact}, \ref{section2exact} and \ref{section3exact}, and the corresponding master equations, we must
make a careful choice of a number numerical parameters. In practice, however, this is not a difficult issue. The essential ingredient is to choose the couplings to the bath according to the desired spectral density. Throughout this paper, we have made the choice (\ref{Ohmic}),
\[
J(\omega)=\sum_jg_j^2\delta(\omega_j-\omega)\approx\alpha\omega e^{-\omega/\omega_c}.
\]
The first step in picking $g_j$ is to remove the Dirac delta functions by integrating over a frequency range bounded by a frequency cut-off $\omega_{\mathrm{max}}$,
\[
\sum_jg_{j}^2\approx\alpha\int_{0}^{\omega_{\mathrm{max}}}\omega e^{-\omega/2\omega_c}d\omega,
\]
which means
\[
g_{j'}^2\approx\alpha\omega_{j'} e^{-\omega_{j'}/2\omega_c}\Delta\omega_{j'}
\]
due to the decomposition of the integral in terms of Riemann sums. We should also take care to set the range of oscillators, $\omega_{\mathrm{max}}$, large enough to cover (\ref{Ohmic}) significantly. For example, if we take $\omega_1=c$, with $c$ small, then one possible convention is to take $\omega_{\mathrm{max}}$ such that $J(\omega_{\mathrm{max}})= J(c)$, and so we neglect all possible oscillators with coupling constant less than $\sqrt{J(c)\Delta\omega_1}$. Another polisher convention is to take $\omega_1$ and $\omega_{\mathrm{max}}$ such that
\begin{eqnarray*}
\fl \int_0^{\omega_1}J(\omega)d\omega=\int_{\omega_{\mathrm{max}}}^\infty J(\omega)d\omega \\ \Rightarrow\omega_c-e^{-\omega_1/\omega_c}(\omega_1+\omega_c)=(\omega_{\mathrm{max}}+\omega_c)e^{-\omega_{\mathrm{max}}/\omega_c}.
\end{eqnarray*}
However, in practice this choice is not really a crucial point.
%--------------------------------------------------------------------------------------------------------------------------------------%
%--------------------------------------------------------------------------------------------------------------------------------------%

\section{Derivation of Markovian Master equations}

\subsection{Two coupled damped harmonic oscillators, small $\beta$} \label{appendixSmallbeta}

We can derive Markovian master equations like (\ref{MasterAprox}) from the microscopic model by the following procedure. The Von Neumann equation in the interaction picture with respect to the free Hamiltonian $H_{\mathrm{free}}=H_{01}+H_{02}+H_{B1}+H_{B2}$ is
\begin{eqnarray}\label{VonN}
\fl \frac{d}{dt}\tilde{\rho}(t)=-i\beta[\tilde{V}_{12}(t),\tilde{\rho}(t)]-i\alpha[\tilde{V}_{SB}(t),\tilde{\rho}(t)]\nonumber\\
\equiv\beta \mathcal{V}_{12}(t)\tilde{\rho}(t)+\alpha\mathcal{V}_{SB}(t)\tilde{\rho}(t),
\end{eqnarray}
where $\tilde{V}_{SB}(t)=\tilde{V}_{1B1}(t)+\tilde{V}_{2B2}(t)$ and for simplicity we have assumed that the strength of the coupling to each bath is identical (the reader will note afterwards that this is not a crucial assumption). We now define the projector $\mathcal{P}\rho(t)=\Tr_{(B1,B2)}[\rho(t)]\otimes\rho_{\mathrm{th}1}\otimes\rho_{\mathrm{th}2}$, along with $\mathcal{Q}=\mathds{1}-\mathcal{P}$. The application of the projection operators on (\ref{VonN}) yields
\begin{eqnarray}
\frac{d}{dt}\mathcal{P}\tilde{\rho}(t)&=\beta\mathcal{P}\mathcal{V}_{12}(t)\tilde{\rho}(t)+\alpha\mathcal{P}\mathcal{V}_{SB}(t)\tilde{\rho}(t),\label{projectorP2ambientes}\\
\frac{d}{dt}\mathcal{Q}\tilde{\rho}(t)&=\beta\mathcal{Q}\mathcal{V}_{12}(t)\tilde{\rho}(t)+\alpha\mathcal{Q}\mathcal{V}_{SB}(t)\tilde{\rho}(t),
\end{eqnarray}
and so (c.f. section \ref{sectionmaster}) we find a formal solution to the
second equation as
\begin{eqnarray}\label{projectorQ2ambientes}
\fl \mathcal{Q}\tilde{\rho}(t)=\mathcal{G}(t,t_0)\mathcal{Q}\tilde{\rho}(t_0)+\beta\int_{t_0}^tds\mathcal{G}(t,s)\mathcal{Q}\mathcal{V}_{12}(s)\mathcal{P}\tilde{\rho}(s)\nonumber\\
+\alpha\int_{t_0}^tds\mathcal{G}(t,s)\mathcal{Q}\mathcal{V}_{SB}(s)\mathcal{P}\tilde{\rho}(s),
\end{eqnarray}
where
\[
\mathcal{G}(t,s)=\mathcal{T}e^{\int_s^tdt'\mathcal{Q}[\beta\mathcal{V}_{12}(t')+\alpha\mathcal{V}_{SB}(t')]}.
\]
Now the procedure is as follows, we introduce the identity $\mathds{1}=\mathcal{P}+\mathcal{Q}$ in the second term of equation (\ref{projectorP2ambientes}),
\[
\frac{d}{dt}\mathcal{P}\tilde{\rho}(t)=\beta\mathcal{P}\mathcal{V}_{12}(t)\tilde{\rho}(t)+\alpha\mathcal{P}\mathcal{V}_{SB}(t)\mathcal{P}\tilde{\rho}(t)+\alpha\mathcal{P}\mathcal{V}_{SB}(t)\mathcal{Q}\tilde{\rho}(t),
\]
and insert the formal solution (\ref{projectorQ2ambientes}) into the last
term. Recalling the condition (\ref{fuera1termino}) $\mathcal{P}\mathcal{V}\mathcal{P}=0$
and again assuming an initial factorized state ($\mathcal{Q}\rho(t_0)=0$) we find
\[
\frac{d}{dt}\mathcal{P}\tilde{\rho}(t)=\beta\mathcal{P}\mathcal{V}_{12}(t)\tilde{\rho}(t)+\int_{t_0}^tds\mathcal{K}_1(t,s)\mathcal{P}\tilde{\rho}(s)+\int_{t_0}^tds\mathcal{K}_2(t,s)\mathcal{P}\tilde{\rho}(s),
\]
where here the kernels are
\begin{eqnarray}
\mathcal{K}_1(t,s)&=&\alpha\beta\mathcal{P}\mathcal{V}_{SB}(t)\mathcal{G}(t,s)\mathcal{Q}\mathcal{V}_{12}(s)\mathcal{P}=0,\nonumber\\
\mathcal{K}_2(t,s)&=&\alpha^2\mathcal{P}\mathcal{V}_{SB}(t)\mathcal{G}(t,s)\mathcal{Q}\mathcal{V}_{SB}(s)\mathcal{P}.
\nonumber
\end{eqnarray}
The first vanishing because $\mathcal{V}_{12}(s)$ commutes with $\mathcal{P}$
and $\mathcal{Q}\mathcal{P}=0$. If we consider the second kernel, weak coupling
implies $\alpha\gtrsim\beta$, and so to second order in  $\alpha$ and $\beta$ this becomes
\[
\mathcal{K}_2(t,s)=\alpha^2\mathcal{P}\mathcal{V}_{SB}(t)\mathcal{Q}\mathcal{V}_{SB}(s)\mathcal{P}+\mathcal{O}(\alpha^3,\alpha^2\beta),
\]
which has exactly the same form as (\ref{kernel2ndorder}) and therefore the equation of motion becomes
\[
\frac{d}{dt}\mathcal{P}\tilde{\rho}(t)=\beta\mathcal{P}\mathcal{V}_{12}(t)\tilde{\rho}(t)+\alpha^2\int_{t_0}^tds\mathcal{P}\mathcal{V}_{SB}(t)\mathcal{V}_{SB}(s)\mathcal{P}\tilde{\rho}(s).
\]
Finally we note that
\begin{eqnarray*}
\fl \Tr_{B1,B2}\left[\tilde{V}_{1B1}(t)\tilde{V}_{2B2}(t')\left(\rho_{\mathrm{th}1}\otimes\rho_{\mathrm{th}2}\right)\right]=\\
\Tr_{B1}[\tilde{V}_{1B1}(t)\rho_{\mathrm{th}1}]\Tr_{B2}[\tilde{V}_{2B2}(t')\rho_{\mathrm{th}2}]=0,
\end{eqnarray*}
because our interactions individually hold $\Tr_{B1}[\tilde{V}_{1B1}\rho_{\mathrm{th}1}]=\Tr_{B2}[\tilde{V}_{2B2}\rho_{\mathrm{th}2}]=0$, so $\mathcal{P}\mathcal{V}_{1B1}\mathcal{V}_{2B2}\mathcal{P}=\mathcal{P}\mathcal{V}_{2B2}\mathcal{V}_{1B1}\mathcal{P}=0$ and then
\begin{eqnarray*}
\fl \frac{d}{dt}\mathcal{P}\tilde{\rho}(t)=\beta\mathcal{P}\mathcal{V}_{12}(t)\tilde{\rho}(t)+\alpha^2\int_{t_0}^tds\mathcal{P}\mathcal{V}_{1B1}(t)\mathcal{V}_{1B1}(s)\mathcal{P}\tilde{\rho}(s)\\
+\alpha^2\int_{t_0}^tds\mathcal{P}\mathcal{V}_{2B2}(t)\mathcal{V}_{2B2}(s)\mathcal{P}\tilde{\rho}(s),
\end{eqnarray*}
which may be rewritten as
\begin{eqnarray}
\fl \frac{d}{dt}\tilde{\rho}_S(t)=-i[\tilde{V}_{12}(t),\tilde{\rho}_S(t)]-\int^t_{t_0}dt'\mathrm{Tr_{B1}}[\tilde{V}_{1B1}(t),[\tilde{V}_{1B1}(t'),[\tilde{\rho}_S(t')\otimes\tilde{\rho}_{\mathrm{th}1}(t')]]\nonumber\\
-\int^t_{t_0}dt'\mathrm{Tr_{B2}}[\tilde{V}_{2B2}(t),[\tilde{V}_{2B2}(t'),[\tilde{\rho}_S(t')\otimes\tilde{\rho}_{\mathrm{th}2}(t')]].
\end{eqnarray}
The last quantity in the above equation is just a sum of the individual terms for each bath, which lead, under the standard procedure of section \ref{sectionmaster}, to the (interaction picture) local dissipators $\mathcal{D}_1$ and $\mathcal{D}_2$ and shifts of (\ref{MasterAprox}).

\subsection{Two coupled damped harmonic oscillators, large $\beta$} \label{appendixLargebeta}

First, let us write the Hamiltonian of the two oscillator system in a more convenient way
\[
H_{12}=H_{01}+H_{02}+V_{12}=(a^\dagger_1,a^\dagger_2)\left(
\begin{array}{cc}
\Omega_1 & \beta\\
\beta  & \Omega_2
\end{array}\right)\left(\begin{array}{c}
a_1\\
a_2
\end{array}\right).
\]
We can diagonalize this quadratic form by means of a rotation to get
\[
H_{12}=\Omega_{+}b^\dagger_1 b_1+\Omega_{-}b^\dagger_2 b_2,
\]
where
\[
\Omega_\pm=\frac{(\Omega_1+\Omega_2)\pm\sqrt{4\beta^2+(\Omega_1-\Omega_2)^2}}{2},
\]
and the creation and annihilation operators in the rotated frame are given by
\begin{eqnarray}
b_1&=&a_1\cos(\alpha)-a_2\sin(\alpha)\nonumber, \\
b_2&=&a_1\sin(\alpha)+a_2\cos(\alpha)\nonumber,
\end{eqnarray}
with the angle specified by
\[
\tan(\alpha)=\frac{2\beta}{(\Omega_1-\Omega_2)-\sqrt{4\beta^2+(\Omega_1-\Omega_2)^2}}.
\]
The new operators satisfy the standard bosonic commutation rules $[b_i,b_j^\dagger]=\delta_{ij}$, and so this is nothing more than the decomposition of an oscillatory system in normal modes. For simplicity, let us now take $\Omega_1=\Omega_2=\Omega$, and so
\[
\Omega_\pm=\Omega\pm\beta, \quad \left\{\begin{array}{l}
b_1=\frac{1}{\sqrt{2}}(a_1+a_2)\\
b_2=\frac{1}{\sqrt{2}}(a_1-a_2)
\end{array}\right.,
\]
note that RWA approximation implies $\Omega\gg\beta$ so both normal mode frequencies are positive.

We can reexpress the interactions with the baths in terms of these new operators,
\begin{eqnarray}
V_{1B1}&=&\sum_{j=1}^M\frac{g_{1j}}{\sqrt{2}}[(b_1^\dagger+b_2^\dagger) a_{1j}+(b_1+b_2)a^\dagger_{1j}],\nonumber \\
V_{2B2}&=&\sum_{j=1}^M\frac{g_{2j}}{\sqrt{2}}[(b_1^\dagger-b_2^\dagger) a_{2j}+(b_1-b_2)a^\dagger_{2j}],\nonumber
\end{eqnarray}
the benefit of this is that it allows us to easily deal with the interaction picture with respect to $H_0=H_{12}+H_{B1}+H_{B2}$. By following the method of section \ref{sectionmaster} we obtain the analog of (\ref{RedfieldMarkov}),
\begin{eqnarray}\label{redfield1}
\fl \frac{d}{dt}\tilde{\rho}_S(t)=-\int_0^\infty dt'\Tr_{B1}[\tilde{V}_{1B1}(t),[\tilde{V}_{1B1}(t-s),\tilde{\rho}_S(t)\otimes\rho_{\mathrm{th}1}]]\nonumber\\
-\int_0^\infty ds\Tr_{B2}[\tilde{V}_{2B2}(t),[\tilde{V}_{2B2}(t-s),\tilde{\rho}_S(t)\otimes\rho_{\mathrm{th}2}]],
\end{eqnarray}
where we have noted $\mathcal{P}\mathcal{V}_{1B1}\mathcal{V}_{2B2}\mathcal{P}=\mathcal{P}\mathcal{V}_{2B2}\mathcal{V}_{1B1}\mathcal{P}=0$. Each of the above terms correspond, essentially, to one of a pair of two free harmonic oscillators with frequencies $\Omega_+$ and $\Omega_-$, coupled to a common bath. Consequently, we can deal with them separately. Starting with the first term
\begin{equation}\label{L1}
\mathcal{L}_1(\tilde{\rho}_S)=-\int_0^tds\Tr_{B1}[\tilde{V}_{1B1}(t),[\tilde{V}_{1B1}(t-s),\tilde{\rho}_S(t-s)\otimes\rho_{\mathrm{th}1}]],
\end{equation}
we decompose the interaction in to eigenoperators of $[H_{12},\cdot]$ (see (\ref{eigenoperators}))
\begin{equation}\label{1B1desc}
V_{1B1}=\sum_{k}A_{k}\otimes B_{k},
\end{equation}
with
\begin{eqnarray}
A_1=\frac{1}{\sqrt{2}}(b_1+b_2), \quad A_2=\frac{1}{\sqrt{2}}(b_1^\dagger+b_2^\dagger),\nonumber\\
B_1=\sum_{j=1}^Mg_{1j}a_{1j}^\dagger, \quad B_2=\sum_{j=1}^Mg_{1j}a_{1j}.
\end{eqnarray}
Notice the $A_1$ operator can be written as $A_1=A_1(\Omega_+)+A_1(\Omega_-)$, where $A_1(\Omega_+)=b_1/\sqrt{2}$ and $A_1(\Omega_-)=b_2/\sqrt{2}$ are already the eigenoperators of $[H_{12},\cdot]$ with eigenvalues $-\Omega_+$ and $-\Omega_-$ respectively. Similarly $A_2=A_2(-\Omega_+)+A_2(-\Omega_-)$, with $A_2(-\Omega_+)=b^\dagger_1/\sqrt{2}$ and $A_2(-\Omega_-)=b^\dagger_2/\sqrt{2}$, and so we can write (\ref{1B1desc}) as
\begin{equation}\label{1B1eigendesc}
V_{1B1}=\sum_{k}A_{k}\otimes B_{k}=\sum_{\nu,k} A_k(\nu)\otimes B_k=\sum_{\nu,k} A_k^\dagger(\nu)\otimes B^\dagger_k,
\end{equation}
which in interaction picture becomes
\[
\tilde{V}_{1B1}(t)=\sum_{\nu,k} e^{-i\nu t}A_k(\nu)\otimes \tilde{B}_k(t)=\sum_{\nu,k} e^{i\nu t}A_k^\dagger(\nu)\otimes \tilde{B}^\dagger_k(t).
\]
Now, for the first element of (\ref{leftFourier}) we have
\begin{eqnarray}
\Gamma_{1,1}(\nu)&=&\sum_{j,j'}g_{1j}g_{1j'}\int_0^\infty ds e^{i(\nu-\omega_{1j}) s}\Tr\left(\rho_{B1}^\mathrm{th}a_{1j}a^\dagger_{1j'}\right)\nonumber\\
&=&\sum_{j=1}^Mg_{1j}^2\int_0^\infty ds e^{i(\nu-\omega_{1j}) s}[\bar{n}_1(\omega_{1j})+1],
\end{eqnarray}
where the mean number of quanta in the first bath $\bar{n}_1(\omega_{1j})$ with frequency $\omega_{1j}$, is given by the Bose-Einstein distribution (\ref{n1}). Going to the continuous limit we take $M\rightarrow\infty$ and introduce the spectral density of the first bath $J_{1}(\omega)=\sum_jg_{1j}^2\delta(\omega-\omega_{1j})$,
\[
\Gamma_{1,1}(\nu)=\int_0^\infty d\omega J_{1}(\omega)\int_0^\infty ds e^{i(\nu-\omega) s}[\bar{n}_{1}(\omega)+1].
\]
Now using the well-know formula from distribution theory,
\[
\int_0^\infty dx e^{ixy}=\pi\delta(y)+i\mathrm{P.V.}\left(\frac{1}{y}\right),
\]
and assuming $\nu>0$, we split into real and imaginary parts,
\[
\Gamma_{1,1}(\nu)=\gamma_1(\nu)[\bar{n}_1(\nu)+1]+i[\Delta_1(\nu)+\Delta'_1(\nu)],
\]
where
\begin{eqnarray}
\gamma_1(\nu)&=&\pi J_1(\nu),\nonumber\\
\Delta_1(\nu)&=&\mathrm{P.V.}\int^\infty_0d\omega\frac{J_1(\omega)}{\nu-\omega},\nonumber\\
\Delta'_1(\nu)&=&\mathrm{P.V.}\int^\infty_0d\omega \frac{J_1(\omega)\bar{n}_1(\omega)}{\nu-\omega}.
\end{eqnarray}
Similar calculations give ($\nu>0$)
\begin{eqnarray}
\Gamma_{1,2}(-\nu)&=&\Gamma_{2,1}(\nu)=0,\\
\Gamma_{2,2}(-\nu)&=&\gamma_1(\nu)\bar{n}_1(\nu)-i\Delta'_1(\nu).
\end{eqnarray}
Thus, equation (\ref{L1}) becomes
\begin{eqnarray}\label{masternosecular}
\fl \mathcal{L}_1(\tilde{\rho}_S)=\sum_{\nu,\nu'}e^{i(\nu'-\nu)t}\Gamma_{1,1}(\nu)[A_1(\nu)\tilde{\rho}_S(t),A_1^\dagger(\nu')]\nonumber\\
+e^{i(\nu-\nu')t}\Gamma_{1,1}^\ast(\nu)[A_1(\nu'),\tilde{\rho}_S(t)A_1^\dagger(\nu)]\nonumber\\
+e^{i(\nu'-\nu)t}\Gamma_{2,2}(\nu)[A_2(\nu)\tilde{\rho}_S(t),A_2^\dagger(\nu')]\nonumber\\
+e^{i(\nu-\nu')t}\Gamma_{2,2}^\ast(\nu)[A_2(\nu'),\tilde{\rho}_S(t)A_2^\dagger(\nu)].
\end{eqnarray}
Next we perform the secular approximation; the cross terms $\nu'\neq\nu$ in the above expression, which go as $e^{\pm2\beta t i}$, can be neglected provided that $2\beta$ is large in comparison with the inverse of the relaxation rate $(\beta\gg\alpha)$ and so we obtain
\begin{eqnarray}\label{L12}
\fl \mathcal{L}_1(\tilde{\rho}_S)=-i\frac{\Delta_1(\Omega_+)}{2}[b_1^\dagger b_1,\tilde{\rho}_S(t)]-i\frac{\Delta_1(\Omega_-)}{2}[b_2^\dagger b_2,\tilde{\rho}_S(t)]\nonumber\\
+\gamma_{1}(\Omega_+)[\bar{n}_1(\Omega_+)+1]\left(b_1\tilde{\rho}_S(t)b_1^\dagger-\frac{1}{2}\{b_1^\dagger b_1,\tilde{\rho}_S(t)\}\right)\nonumber\\
+\gamma_{1}(\Omega_+)\bar{n}_1(\Omega_+)\left(b_1^\dagger\tilde{\rho}_S(t)b_1-\frac{1}{2}\{b_1 b_1^\dagger,\tilde{\rho}_S(t)\}\right)\nonumber\\
+\gamma_{1}(\Omega_-)[\bar{n}_1(\Omega_-)+1]\left(b_2\tilde{\rho}_S(t)b_2^\dagger-\frac{1}{2}\{b_2^\dagger b_2,\tilde{\rho}_S(t)\}\right)\nonumber\\
+\gamma_{1}(\Omega_-)\bar{n}_1(\Omega_-)\left(b_2^\dagger\tilde{\rho}_S(t)b_2-\frac{1}{2}\{b_2 b_2^\dagger,\tilde{\rho}_S(t)\}\right)\nonumber.
\end{eqnarray}

Returning to equation (\ref{redfield1}), for the second term,
\[
\mathcal{L}_2(\tilde{\rho}_S)=-\int_0^\infty ds\Tr_{B1}[\tilde{V}_{2B2}(t),[\tilde{V}_{2B2}(t-s),\tilde{\rho}_S(t)\otimes\rho_{\mathrm{th}2}]],
\]
the situation is essentially the same, since the minus sign in $b_2$ only modifies the cross terms, which we neglect in the secular approximation. Following similar steps as in the above we obtain the same form (\ref{L12}) for $\mathcal{L}_2$, with the replacements $\gamma_1\rightarrow\gamma_2$, $\Delta_1\rightarrow\Delta_2$ and $\bar{n}_1\rightarrow\bar{n}_2$, where the subscript 2 refers to the corresponding expression with the spectral density and temperature of the second bath. Therefore putting together both quantities, and returning to the Schr\"odinger picture
\begin{eqnarray}
\fl \frac{d}{dt}\rho_S(t)=-i\left[\Omega_1+\Delta_1(\Omega_+)/2+\Delta_2(\Omega_+)/2\right][b_1^\dagger b_1,\rho_S(t)]\nonumber\\
-i\left[\Omega_2+\Delta_1(\Omega_-)/2+\Delta_2(\Omega_-)/2\right][b_2^\dagger b_2,\rho_S(t)]\nonumber\\
+\{\gamma_1(\Omega_+)[\bar{n}_1(\Omega_+)+1]+\gamma_2(\Omega_+)[\bar{n}_2(\Omega_+)+1]\}\left(b_1\rho_S(t)b_1^\dagger-\frac{1}{2}\{b_1^\dagger b_1,\rho_S(t)\}\right)\nonumber\\
+[\gamma_1(\Omega_+)\bar{n}_1(\Omega_+)+\gamma_2(\Omega_+)\bar{n}_2(\Omega_+)]\left(b_1^\dagger\rho_S(t)b_1-\frac{1}{2}\{b_1 b_1^\dagger,\rho_S(t)\}\right)\nonumber\\
+\{\gamma_1(\Omega_-)[\bar{n}_1(\Omega_-)+1]+\gamma_2(\Omega_-)[\bar{n}_2(\Omega_-)+1]\}\left(b_2\rho_S(t)b_2^\dagger-\frac{1}{2}\{b_2^\dagger b_2,\rho_S(t)\}\right)\nonumber\\
+[\gamma_1(\Omega_-)\bar{n}_1(\Omega_-)+\gamma_2(\Omega_-)\bar{n}_2(\Omega_-)]\left(b_2^\dagger\rho_S(t)b_2-\frac{1}{2}\{b_2 b_2^\dagger,\rho_S(t)\}\right).
\end{eqnarray}
It is manifestly clear that this equation is of the Kossakowski-Lindblad form. Finally, we rewrite the operators $b_1$ and $b_2$ in terms of $a_1$ and $a_2$ to arrive at equation (\ref{MasterLbeta}).

It is worth mentioning that similar equations for coupled harmonic oscillators have been given previously (see for example \cite{CarmichaelWals,dePonte}), but not in the Kossakowski-Lindblad form, since in those derivations the secular approximation is not taken.

\subsection{Driven damped harmonic oscillator}\label{appendixDriven}

To derive a completely positive Markovian master equation valid for large Rabi frequencies $r$ we must work in the interaction picture generated by the unitary propagator $U(t_1,t_0)=\mathcal{T}e^{-i\int_{t_0}^{t_1}H_1(t')dt'}$, where
\begin{equation}
H_1(t)=\Omega a^\dagger a+r(a^\dagger e^{-i\omega_Lt}+a e^{i\omega_Lt})+\sum_{j=1}^M\omega_ja^\dagger_j a_j.
\end{equation}
Taking $t_0=0$ without lost of generality, the time-evolution equation for $\tilde{\rho}(t)=U^\dagger(t,0)\rho(t)U(t,0)$ is
\begin{equation}\label{ecPictureRara}
\dot{\tilde{\rho}}(t)=-i[\tilde{V}(t),\tilde{\rho}(t)],
\end{equation}
so by following the analogous procedure for time-independent generators, one immediately deals with the problem that is not clear whether there exists a similar eigenoperator decomposition for $\tilde{V}(t)=U^\dagger(t,0)VU(t,0)$ ($V=\sum_{j=1}^Mg_j(a^\dagger a_j + a a^\dagger_j)$) as in (\ref{eigenoperatorsDesc}) and (\ref{eigenoperators}). Note however that the operator $\tilde{A}_1(t)=\tilde{a}(t)$ satisfies a differential equation with periodic terms
\begin{equation}\label{a(t)diff}
i\dot{\tilde{a}}(t)=[\tilde{a}(t),H_0(t)]=\Omega \tilde{a}(t)+re^{-i\omega_Lt}.
\end{equation}
This kind of equation can be studied with the well-established Floquet theory (see for example \cite{Chicone,Ince}), particularly it is possible to predict if its solution is a periodic function. In such a case, the operator in the new picture would have a formal decomposition similar to that in (\ref{eigenoperatorsDesc}) and (\ref{eigenoperators}), such that $\tilde{A}_k(t)=\sum_\nu A_k(\nu)e^{i\nu t}$. This would then allow us to follow a similar procedure to that for time-independent Hamiltonians. Note that the importance of such a decomposition is that the operators $A_k(\nu)$ are themselves time-independent. Such ideas have already been used before in, for instance, \cite{BrPe97,Hanggi99}.

The solution to equation (\ref{a(t)diff}), with the initial condition $\tilde{a}(0)=a$ and for $\Omega\neq\omega_L$ is given by
\begin{equation}\label{a(t)off-r}
\tilde{a}(t)=\frac{r(e^{-i\omega_Lt}-e^{-i\Omega t})+a(\omega_L-\Omega)e^{-i\Omega t}}{\omega_L-\Omega},
\end{equation}
so in this case the solution is periodic and the desired decomposition $\tilde{A}_1(t)=\sum_\nu A_1(\nu)e^{i\nu t}$ is
\[
\tilde{A}_1(t)=A_1(\omega_L)e^{-i\omega_Lt}+A_2(\Omega)e^{-i\Omega t},
\]
where $A_1(\omega_L)=\frac{r}{\omega_L-\Omega}\id$ and $A_1(\Omega)=a-\frac{r}{\omega_L-\Omega}\id=a-A_1(\omega_L)$. Similarly
\[
\tilde{A}_2(t)=A_2(-\omega_L)e^{i\omega_Lt}+A_2(-\Omega)e^{i\Omega t},
\]
with $A_2(-\omega_L)=\frac{r}{\omega_L-\Omega}\id=A_1(\omega_L)$ and $A_2(-\Omega)=a^\dagger-\frac{r}{\omega_L-\Omega}\id=a-A_2(-\omega_L)$. Thus we get an equation analogous to (\ref{masternosecular}), where the coefficients are:
\begin{eqnarray*}
\Gamma_{11}(\nu)&=&\gamma(\nu)[\bar{n}(\nu)+1]+i[\Delta(\nu)+\Delta'(\nu)],\quad(\nu>0)\\
\Gamma_{12}(\nu)&=&\Gamma_{21}(\nu)=0,\\
\Gamma_{22}(\nu)&=&\gamma(-\nu)\bar{n}(-\nu)-i\Delta'(-\nu)\quad(\nu<0).
\end{eqnarray*}
Before continuing note that in the perturbative series of (\ref{ecPictureRara}), the ``strength'' of the interaction $\tilde{V}(t)$ is now not solely dependent on the coupling with the bath. This is because the operators $A(\nu)$ depend linearly on $\frac{r}{w^L-\Omega}$, so when this ratio becomes large we expect that the approximation breaks down, i.e. for $r\gg 1 $ or very close to resonance $|w^L-\Omega|\approx0$.

Next we assume that the detunning is large enough $|\omega_L-\Omega|\gg\alpha$, $|\omega_L-\Omega|^2\gg\alpha r$ in order to make the secular approximation and after some tedious, but straightforward, algebra we find the master equation in the interaction picture to be
\begin{eqnarray}
\fl \frac{d}{dt}\tilde{\rho}_S=-i[\Delta(\Omega)a^\dagger a -\frac{\Delta(\Omega)r}{\omega_L-\Omega}(a+a^\dagger)\nonumber\\
+\frac{\gamma(\Omega)r}{\omega_L-\Omega}\frac{a-a^\dagger}{i},\tilde{\rho}_S]+D(\tilde{\rho}_S),
\end{eqnarray}
where $\mathcal{D}(\cdot)$ has again the form of (\ref{Dissipator}). Finally, on returning to the Schr\"odinger picture we have,
\begin{eqnarray}\label{masterstrongfield}
\fl \frac{d}{dt}\rho_S=-i[H_1(t),\rho_S]+U(t,0)\dot{\tilde{\rho}}_SU^\dagger(t,0)\nonumber\\
=-i[\bar{\Omega}a^\dagger a+\bar{r}e^{i\omega_Lt}a+\bar{r}^{\ast}e^{-i\omega_Lt}a^\dagger,\rho_S]+D(\rho_S),
\end{eqnarray}
where $\bar{\Omega}=\Omega+\Delta(\Omega)$ and
\begin{equation}
\bar{r}=r\left[1+\frac{\Delta(\Omega)+i\gamma(\Omega)}{\Omega-\omega_L}\right].
\end{equation}
So in this master equation the Rabi frequency is renormalized by the effect of the bath. It is worth noting that at first order in $r$ and the coupling $\alpha$ we obtain equation (\ref{rverysmall}). This is as expected, given the arguments in section \ref{sectionmasteraprox}.

For an arbitrary driving frequency a Markovian master equation is difficult to obtain as we cannot, in general, make the secular approximation (apart from the perturbative condition $|w^L-\Omega|\nsim0$). This can be illustrated in the extreme case of resonance $\omega_L=\Omega$. Solving equation (\ref{a(t)diff}) under this condition we find
\begin{equation}\label{a(t)res}
\tilde{a}(t)=e^{-i\Omega t}(a-irt),
\end{equation}
and so one can see that $\tilde{a}(t)$ is not a periodic function, so the desired decomposition as a sum of exponentials with time-independent coefficients does not exist. On the other hand, the decomposition (\ref{a(t)off-r}) tends to (\ref{a(t)res}) in the limit $\omega_L\rightarrow\Omega$, so we may attempt to work with this decomposition and wonder whether on resonance the new master equation holds in this limit as well (in fact, we have shown that this is not true in section \ref{DrivenDampedHOSimulation}). The only problem to deal with is the possible lack of positivity due to the absence of the secular approximation. However, note that in this particular case only a commutator term arises from the cross terms in the analog of equation (\ref{masternosecular}), so positivity is not lost. In fact, we obtain an equation similar to (\ref{masterstrongfield}) except for an additional correction to the Rabi frequency:
\begin{equation}
\bar{r}=r\left[1+\frac{\Delta(\Omega)+i\gamma(\Omega)}{\Omega-\omega_L}-\frac{\Delta(\omega_L)+i\gamma(\omega_L)}{\Omega-\omega_L}\right].
\end{equation}
Note that to first order in $r$ and $\alpha$ we again obtain the equation (\ref{MasterFree}).


\section*{References}
\begin{thebibliography}{10}
\bibitem{general} Peier W 1972 {\it Physica} \textbf{57} 565; Shibata F and Hashitsume N 1974 {\it Z. Phys.} \textbf{B34} 197; Willis C R and Picard R H 1974 {\it Phys. Rev. A} \textbf{9} 1343; Schwendimann P 1977 {\it Z. Phys.} \textbf{B26} 63.
%
\bibitem{RevKoss} Gorini V, Frigerio A, Verri M, Kossakowski A and Sudarshan E C G 1978 {\it Rep. Math. Phys.} \textbf{13} 149--73.
%
\bibitem{BreuerPetruccione} Breuer H-P and Petruccione F 2002 {\it The Theory of Open
Quantum Systems} (New York: Oxford University
Press).
%
\bibitem{Davies1} Davies E B 1974 {\it Comm. Math. Phys.} \textbf{39} 91--110.
%
\bibitem{Davies2} Davies E B 1976 {\it Math. Ann.} \textbf{219} 147--158.
%
\bibitem{Koss-Lind} Gorini V, Kossakowski A and Sudarshan E C G 1976 {\it J. Math.
Phys.} \textbf{17} 821; Lindblad G 1976 {\it Commun. Math. Phys.}
\textbf{48} 119.
%
\bibitem{EisertPlenio03} Eisert J, Plenio M B 2003 {\it Int. J. Quant. Inf.} \textbf{1} 479.
%
\bibitem{Spohn} Dumcke R and Spohn H 1979 {\it Z. Phys.} \textbf{B34} 419.
%
\bibitem{giovana} Englert B-G and Morigi G 2002 {\it Coherent Evolution in Noisy Environments (Lecture Notes in Physics)}
vol 611 (Berlin: Springer) p 55.
%
\bibitem{ModernCohen} Kryszewski S and Czechowska-Kryszk J 2008 Master equation - tutorial approach {\it Preprint} quant-ph/08011757.
%
\bibitem{wolf} Wolf M M, Eisert J, Cubitt T S and Cirac J I 2008 {\it Phys. Rev.
Lett.} \textbf{101} 150402; Wolf M M and Cirac J I 2008 {\it Comm. Math. Phys.} \textbf{279} 147.

\bibitem{breuer} Breuer H-P, Laine E-M and Piilo J 2009 {\it Phys. Rev. Lett.} \textbf{103} 210401.
%
\bibitem{rivas} Rivas A, Huelga S F and Plenio M B 2010 {\it Phys. Rev. Lett.} \textbf{105} 050403.
%
\bibitem{resto} Lu X-M, Wang X and Sun C P 2010 {\it Phys. Rev. A} \textbf{82} 042103.
%
\bibitem{Haake} Haake F 1973 {\it Statistical Treatment of Open Systems by Generalized Master Equations (Springer Tracts in
Modern Physics)} vol 66 (Berlin: Springer) pp 98–168.
%
\bibitem{Gardiner} Gardiner C W and Zoller P 2004 {\it Quantum Noise} (Berlin: Springer).
%
\bibitem{Puri} Puri R R 2001 {\it Mathematical Methods of Quantum Optics} (Berlin: Springer).
%
\bibitem{Carmichael} Carmichael H J 1999 {\it Statistical Methods in Quantum Optics I: Master Equations and Fokker-Plack Equations} (Berlin: Springer).
%
\bibitem{Weiss} Weiss U 2008 {\it Quantum Dissipative Systems} (Singapore: World Scientific).
%
\bibitem{Cohen} Cohen-Tannoudji C, Dupont-Roc J and Grynberg G 1992 {\it Atom- Photon Interactions} (New York: John Wiley \& Sons).
%
\bibitem{HOscillator} Puri R R and Lawande S V 1977 {\it Phys. Lett.} \textbf{64A} 143-5; Puri R R and Lawande S V 1978 {\it Phys. Lett.} \textbf{69A} 161-3; Hu B L, Paz J P and Zhang Y 1992 {\it Phys. Rev. D} \textbf{45} 2843-61; Karrlein R and Grabert H 1997 {\it Phys. Rev. E} \textbf{55} 153--64.
    %
\bibitem{recua} Jeong H Lee J and Kim M S 2000 {\it Phys. Rev. A} \textbf{61} 052101;
Lee J, Kim M S and Jeong H 2000 {\it Phys. Rev. A} \textbf{62} 032305;
Kim M S and Lee J 2002 {\it Phys. Rev. A} \textbf{66} 030301R;
Scheel S and Welsch D -G 2001 {\it Phys. Rev. A} \textbf{64} 063811;
Hiroshima T 2001 {\it Phys. Rev. A} \textbf{63} 022305; Prauzner-Bechcicki J S 2004 {\it J. Phys. A: Math. Gen.} \textbf{37} L173; An J -H and Zhang W -M 2007 {\it Phys. Rev. A} \textbf{76} 042127; Paz J P and Roncaglia A 2009 {\it Phys. Rev. Lett.} \textbf{100} 220401; Paz J P and Roncaglia A 2009 {\it Phys. Rev. A} \textbf{79} 032102.
%
\bibitem{hu2} Chou C -H, Yu T and Hu B -L 2008 {\it Phys. Rev. E} \textbf{77} 011112.
%
\bibitem{Nakajima} Nakajima S 1958 {\it Progr. Theor. Phys.} {\bf 20} 984.
%
\bibitem{Zwanzig} Zwanzig R 1960 {\it J. Chem. Phys.} {\bf 33} 1338--41.
%
\bibitem{Chicone} Chicone C 2006 {\it Ordinary Differential Equations with Applications} (New York: Springer).
%
\bibitem{Ince} Ince E L 1956 {\it Ordinary Differential Equations} (New York: Dover).
%
\bibitem{reedsimon1} Reed M and Simon B 1980 {\it Methods of Modern Mathematical Physics I} (San Diego: Academic Press).
%
\bibitem{Bures} Bures D 1969 {\it Trans. Am. Math. Soc.} \textbf{135} 199.
%
\bibitem{Scutaru} Scutaru H 1998 {\it J. Phys. A: Math. Gen.} \textbf{31} 3659; Paraoanu Gh -S and Scutaru H 2000 {\it Phys. Rev. A} \textbf{61} 022306.
%
\bibitem{RWA} In fact, note that the rotating wave Hamiltonian is not always an approximation, and there are physical systems which described physically by that Hamiltonian, for instance typically in situations where the total number of excitations is preserved.
%
\bibitem{Alessio} Paris M G A, Illuminati F, Serafini A, and De Siena S 2003 {\it Phys. Rev A} \textbf{68} 012314.
%
\bibitem{zeta} For more details see Apostol T 1976 {\it Introduction to Analytic Number Theory} (New York: Springer).
%
\bibitem{Silbey} On this topic see for example the work Su\'arez A, Silbey R and Oppenheim I 1992 {\it J. Chem. Phys} \textbf{97} 5101--07.
%
\bibitem{Oscillators68} Estes L E, Keil T H and Narducci L M 1968 {\it Phys. Rev.} \textbf{175} 286.
%
\bibitem{ORHF09} Oxtoby N P, Rivas A, Huelga S F and Fazio R 2009 {\it New J. Phys.} \textbf{11} 063028.
%
\bibitem{Ht} Davies E B and Spohn H 1978 {\it J. Stat. Phys.} \textbf{19} 511; Alicki R 1979 {\it J. Phys. A: Math. Gen.} \textbf{12} L103.
%
%\bibitem{dimitris} Henrich M J,  Michel M, Hartmann M, Mahler G and Gemmer J 2005 {\it Phys. Rev. E} \textbf{72}, 026104; Tsomokos D I, Hartmann M J, Huelga S F and Plenio M B 2007 {\it New J. Phys.} \textbf{9} 79; Prosen T and \v{Z}unkovi\v{c} 2010 {\it New J. Phys.} \textbf{12}, 025016.
%
\bibitem{iones} Garg A 1996 {\it Phys. Rev. Lett.}  \textbf{77} 964; Porras D, Marquardt F, von Delft J and Cirac J I 2008 {\it Phys. Rev. A}  \textbf{78} 010101R.
%
\bibitem{scq} Romito A, Fazio R and Bruder C 2005 {\it Phys. Rev. B} \textbf{71} 100501.
%
\bibitem{nv} Gaebel T \etal 2006 {\it Nature Physics} \textbf{2} 408--413.
%
\bibitem{bio} Mohseni M, Rebentrost P, Lloyd S and Aspuru-Guzik A 2008 {\it J. Chem. Phys.} \textbf{129} 174106; Plenio M B and Huelga S F 2008 {\it New J. Phys.} \textbf{10} 113019; Olaya-Castro A, Lee C F, Olsen F F and Johnson N F 2008 {\it Phys. Rev. B} \textbf{78} 085115; Rebentrost P, Mohseni M, Kassal I, Lloyd S and Aspuru-Guzik A 2009 {\it New J. Phys.} \textbf{11} 033003; Caruso F, Chin A W, Datta A, Huelga S F and Plenio M B 2009 {\it J. Chem. Phys.} \textbf{131} 105106; Thorwart M \etal 2009 {\it Chem. Phys. Lett.} \textbf{478} 234.
%
\bibitem{Topological} See for example: Alicki R, Fannes M and Horodecki M 2009 {\it  J. Phys. A: Math. Theor.} \textbf{42} 065303; Bombin H, Chhajlany R W, Horodecki M and Martin-Delgado M A 2009 Self-Correcting Quantum Computers {\it Preprint} quant-ph/09075228; Chesi S, R\"othlisberger B and Loss D 2010 {\it Phys. Rev. A} \textbf{82} 022305.
%
\bibitem{CarmichaelWals} Carmichael H J and Walls D F 1973 {\it J. Phys. A: Math. Nucl. Gen.} \textbf{6} 1552-64.
%
\bibitem{dePonte} de Ponte M A, de Oliveira M C and Moussa M H Y 2004 {\it Phys. Rev. A} \textbf{70} 022324; {\it Phys. Rev. A} \textbf{70} 022325; 2005 {\it Ann. Phys.} \textbf{317} 72.
%
\bibitem{BrPe97} Breuer H P and Petruccione F 1997 {\it Phys. Rev. A} \textbf{55} 3101.
%
\bibitem{Hanggi99} Kohler S, Dittrich T and H\"anggi P 1999 {\it Phys. Rev. E} \textbf{55} 300.
\end{thebibliography}
\end{document}